\documentclass[12pt]{article}
\usepackage[english]{babel}
\usepackage[letterpaper,top=2cm,bottom=2cm,left=2cm,right=2cm,marginparwidth=1.75cm]{geometry}
\usepackage{amsfonts}   
\usepackage{amsmath}
\usepackage{makecell}
\usepackage{multirow}
\usepackage{graphicx}
\usepackage{amssymb}       
\usepackage{algorithm}      
\usepackage{algpseudocode}  

\usepackage[colorlinks=true, allcolors=blue]{hyperref}
\usepackage{comment}
\title{A new wavelet-based variational family with copula dependence structures}
\author{
Giovanni Piccirilli\thanks{Corresponding author. Email: giovanniopcl@gmail.com}
\and
Aluísio Pinheiro
}

\begin{document}
\maketitle

\begin{abstract}
Variational inference (VI) has become a widely used approach for scalable Bayesian inference, but its performance strongly depends on the flexibility of the chosen variational family. In this work, we propose a novel variational family that combines wavelet-based representations for marginal posterior densities with copula functions to model dependence structures. The marginal distributions are constructed using coefficients from the discrete wavelet transform, providing a flexible and adaptive framework capable of capturing complex features such as asymmetry. The joint distribution is then obtained through a copula, allowing for explicit modeling of dependence among parameters, including both independence and Gaussian copula structures. We develop an efficient estimation procedure based on Monte Carlo approximations of the evidence lower bound (ELBO) and automatic differentiation, enabling scalable optimization using gradient-based methods. Through extensive simulation studies, including logistic regression, sparse linear models, and hierarchical models, we demonstrate that the proposed approach achieves posterior mean estimates comparable to Markov chain Monte Carlo (MCMC) methods, while providing improved uncertainty quantification relative to standard variational approaches. Applications to hierarchical logistic regression and Bayesian conditional transformation models further illustrate the practical advantages of the method in complex, high-dimensional settings. The proposed wavelet–copula variational family offers a flexible and computationally efficient alternative for Bayesian inference.
\end{abstract}

\section{Introduction}

Variational inference (VI) is a method from machine learning for approximating probability densities \cite{ref2VB}. Some authors refer to Variational Bayes (VB) when the context of variational application is posterior distributions approximation from some Bayesian model. The VI methods gained popularity in the Bayesian inference due to the massive amount of analyzed data, and the growing need for computationally efficient alternatives to sampling-based methods to make inference in more complex models. In fact, it becomes an alternative for the popular Monte Carlo Markov Chain (MCMC) sampling based methods because it tends to be faster and easier to scale to large-data problems. Some popular R-packages in Bayesian inference, like, Integrated Nested Laplace Approximation (INLA) \cite{rue2009approximate}, or the \texttt{Stan} \cite{StanUserGuide}, are offering alternatives for the users based on VI like VB correction \cite{van2024low}, and the Automatic Variational Differentianion Inference (ADVI) \cite{kucukelbir2017automatic}.

VI is a more scalabe method because turns the sampling from MCMC algorithms, which can be a problem specially in model with a moderate or large number of parameters, into an optimization problem. Besides that, the popularization of Monte Carlo (MC) integration, stochastic optimization algorithms, and the implementation of the Automatic Differentiation (AD) \cite{baydin2018automatic} in many programming language makes the VI more popular for the Bayesian process.

Key references like \cite{ref1VB} and \cite{ref2VB} helped to establish VB as a crucial
tool in probabilistic modeling and machine learning, cementing its role in developing
scalable Bayesian inference techniques. VI algorithms approximate the posterior distribution by a member of a known family of probability densities, called Variational family, which is the one closest to the true posterior terms of Kullback-Leibler Divergence (KLD). There is nothing new about approximation posterior distributions by unknown probability distributions in Bayesian computation. The Laplace approximation \cite{tierney1986accurate}, and ILNA \cite{rue2009approximate} are example of popularized and well established methodologies and produce reasonable approximations. But, in the VI we got a more flexible context by choosing a rich Variational family. The chosen of the Variational family has a huge impact on the goodness of the approximation. \cite{bayes_opt} says that if the true posterior density is included in the variational family the VI select it, as it has the smallest KLD distance. So, it seem obviously that, but it gives the intuition that richer the VI family closer we are to produce better estimates from the true posterior.

However, defining a family is not easy, it should be rich enough but limited at the points such that the search is computationally viable. A popular approach is to parametrize this family in such a way that the inspection for the densities turn out to be the search for parameters of the family, called Variational parameters. The drawback is the fact that we considerably restrain the Variational family when we parametrize it. However, this clarifies the problem since now we are now looking for parameters in a good looking space, like $\mathbb{R}^d$ for some integer $d$ instead of searching for densities. Then, the VI is now an optimization problem, where we need to find the optimal parameter which minimizes the KLD with the aid of some optimization algorithm. In the search for a satisfactory Variational family, \cite{kucukelbir2017automatic} for example, propose the Automatic Differentiation Variational Inference (ADVI), which uses the Gaussian family as variation family where the mean vector and covariance matrix are the variational parameters. About more flexible families, \cite{zobay2014variational} propose an extension to the Gaussian approach which uses Gaussian mixtures as approximations, \cite{tran2015copula} propose copula variational family, \cite{ranganath2016hierarchical} employ a hierarchical Variational Bayes, and \cite{rezende2015variational} use normalizing flows to construct the variational family. 

Wavelet methods provide a flexible and powerful framework for representing complex signals and functions across multiple scales. Unlike classical Fourier-based approaches, which rely on global basis functions, wavelets are localized in both time and frequency, allowing them to capture abrupt changes, local features, and nonstationary behavior. In a statistical context, wavelets have been widely used for dimension reduction and adaptive function approximation (also in density estimation), offering sparse representations that can efficiently capture underlying signal structure.

In this work, we propose a new variational family in which the posterior distribution is constructed by combining marginal wavelet-based densities with a copula that captures the dependence structure among parameters. Within this framework, the posterior marginal densities are parameterized through the coefficients of the Discrete Wavelet Transform (DWT) \cite{nason2008wavelet}, enabling flexible and adaptive marginal approximations. The copula component then induces a joint distribution by introducing additional parameters that explicitly model dependence, with the specific parameterization determined by the chosen copula family.

This works is organized as follows...

\section{Technical Background}
\subsection{Wavelets}

Wavelets are a family of orthonormal basis used to express and approximate functions from the $L_2(\mathbb{R})$ space. They are very popular in time series analysis and signal process specially by their structure extraction of a signal, which includes computationally efficiency of decomposition, and sparsity of coefficients. These ingredients make a good alternative for time series decomposition. However, the use of the wavelets goes beyond this subjects, they are popular in density estimation \cite{pinheiro1997estimating} \cite{donoho1996density}, being an alternative to Kernel estimator highlighted by their computational cost. 

Wavelets are better constructed and understood in a hierarchical framework, the methodology ``breaks" the function in many scales separating tendency and noise. Let us defined the space $V_j$ as the space of functions with details up to some finest scale of resolution. 
Larger values of $j$ correspond to higher resolution levels, meaning that
$V_j$ contains functions with increasingly finer details. Mathematically, this is expressed by the nested structure
\begin{equation}
    V_0 \subset V_1 \subset V_2 \subset \dots .
\end{equation}

As $j$ becomes larger and positive we include more an more functions with finer scales so that $\bigcup V_j =  L_2(\mathbb{R})$. If $f(x)$ is a member of $V_j$, then 

\begin{equation}\label{wave_proper_bases}
    f(2x) \in V_{j + 1} \quad \text{and} \quad f(x-k) \in V_j.
\end{equation}

From $V_0$ constructions, there exists a scaling function $\phi \in V_0$ such that $V_0 = \text{span}\lbrace \phi(x - k), k \in \mathbb{Z} \rbrace$. From (\ref{wave_proper_bases}) $f(2^j x) \in V_j$, so we can scale the $V_0$ base: $\phi_{j,k}(x) = 2^{j/2} \phi(2^jx - k)$, so that, $\lbrace\phi_{j,k}(x) \rbrace$ is a base for $V_j$.

Let $W_j$ be the orthogornal complement of $V_j$ to $V_{j - 1}$, i.e $V_{j + 1} = V_j \oplus W_j$. Given the function $\phi$, one defines the so called mother wavelet as

\begin{equation}
    \psi(x)= \sum_k (-1)^k h_{1 - k} \phi(2x - k)
\end{equation}

\noindent where the coefficients $h_k, k \in \mathbb{N}$ are defined by the following dilation equation

\begin{equation}\label{dilation_equation}
    \phi(x) = \sum_k h_k \phi(2x - k)
\end{equation}

\noindent and, $\lbrace \psi(x - k), k \in \mathbb{Z}\rbrace$ constitutes a base for $W_0$, and scale and translation of this basis defined a basis for $W_j$. Since $\bigcup V_j =  L_2(\mathbb{R})$, and $V_{j + 1} = V_j \oplus W_j$ we say, for some integer $j_0 > 0$, that $ L_2(\mathbb{R}) = V_{j_0} \oplus_{j > j_0} W_j$ which means that any $L_2$ function can be written as 

\begin{equation}\label{functions_wave_l2}
        f(x) = \sum_{k} c_{j_0,k} \phi_{j_0,k}(x) + \sum_{j = j_0}^{\infty}\sum_k d_{j,k}\psi_{j,k}(x)
\end{equation}

\noindent where $c_{j_0,k}$ are called as approximation coefficients for some coarsest level $j_0$, $d_{j,k}$ are the detailed coefficients $\psi$ and $\phi$ are the respectively father and mother wavelets . The expression in (\ref{functions_wave_l2}) is the wavelet representation of $L_2(\mathbb{R})$ functions which contains a smooth part involving the $ \sum_{k} c_{j_0,k} \phi_{j_0,k}(x)$ terms, and a set of details represented by $\sum_{j = j_0}^{\infty}\sum_k d_{j,k}\psi_{j,k}(x)$. 

We will not work with continuous representation of functions $f(x), x \in \mathbb{R}$, but with a discrete representation defined as a set of values $\boldsymbol{f} = (f_1, \dots, f_N)$, that is, a set of values indexed in the integers with some $N \in \mathbb{N}$. In the wavelet context, these values are generally refereed as a signal. What we do in practice is to use the Discrete Wavelet Transformation (DWT), which maps the data from the original domain to the Wavelet domain, to get a discrete representation of the function. The discrete Wavelet representation is feasible due to the so called cascade algorithm which consists of apply a sequence of wavelets filters and downsampling in each step to get a set of approximation and details coefficients \cite{morettin2017wavelets}.

Let $\lbrace h_k, k \in \mathbb{Z} \rbrace$ be the so called low pass filter, where $h_m = 0$ for $m < 0$, or $m \geq L$, where L is an integer which depends on the chosen Wavelet family, and the high pass filter
$\lbrace g_k, k \in \mathbb{Z} \rbrace$, where $g_k = (-1)^kh_{1 -k}$. For a sequence $\boldsymbol{f} = \lbrace f_N \rbrace$ we define the following linear operators: $(Ha)_n  = \sum_k h_{k - 2n}a_k,$ and $(Ga)_n = \sum_k g_{k - 2n}a_k.$

They perform filtering and downsampling (omitting every second entry in the output of filtering), and correspond to a single step in the wavelet decomposition. The algorithm start with the original signal $\boldsymbol{f}$ of length $N = 2^J$ denoted by $\boldsymbol{c}^{(J)}$. At each step we move to a coarser approximation $\boldsymbol{c}^{(j - 1)}$ with $\boldsymbol{c}^{(j - 1)} = H \boldsymbol{c}^{(j)}$ and $\boldsymbol{d}^{(j - 1)} = G \boldsymbol{c}^{(j)}$. The length of $\boldsymbol{c}^{(j - 1)}$ and $\boldsymbol{d}^{(j - 1)}$ are half of the $j$ indexes. The decomposition can be carried out $J$ times, or stopped at the $(J - R)\text{th}$ step that produces

\begin{equation}\label{coef_dwt}
    (\boldsymbol{c}^{(R)}, \boldsymbol{d}^{(R)}, \boldsymbol{d}^{(R+1), \hdots,\boldsymbol{d}^{(J - 1)}})
\end{equation}

\noindent for any $0 < R \leq J - 1$. The reconstruction signal from the coefficients in (\ref{coef_dwt}) are called IDWT. In this process, the operators are $(H^*a)_k = \sum_n h_{k - 2n} a_n$, and $(G^*a)_k = \sum_n g_{k - 2n} a_n $, and the original signal is reconstructed as

\begin{equation}\label{IDWT}
    \boldsymbol{f} = (H^*)\boldsymbol{c}^{(R)} + \sum_{j = R}^{J - 1} (H^*)^jG^* \boldsymbol{d}^{(j)}.
\end{equation}

The DWT gets the original signal and decompose it into coefficients. See that we do not need a direct evaluation of mother and father wavelets because of the (\ref{dilation_equation}) equation, by knowing the filters values of the wavelet, the bases are evaluated iteratively. In time series and process signal it is very useful, specially because these coefficients are sparse, doing some thresholding technique we can discard some of them, and our storage of the signal could be more efficient. The IDWT do just the opposite, it gets the coefficients and reconstruct the signal. If we conserve all coefficients the original signal is reconstructed, but doing some thresholding technique in the coefficients, then we got a approximated signal.


\subsection{Copulas}

A d-dimensional copula, $C:[0,1]^d: \rightarrow [0,1]$ is a cumulative distribution function (CDF) with uniform marginals. Let $\boldsymbol{X} = (X_1, \dots, X_d)$ a multivariate random vector with CDF $F_X$ and with continuous and inreasing marginals. We can expresse the copula $C_X$ associated with $X$ as

\begin{align}\label{copula_cdf}
C_X(u_1, \hdots, u_n) & = P(F_{X_1}(X_1) \leq u_1, \dots, F_{X_d}(X_d) \leq u_d) \nonumber \\
& = P(X_1 \leq F^{-1}_{X_1}(u_1), \dots, X_d \leq F^{-1}_{X_d}(u_d)) \nonumber \\
& = F_{\boldsymbol{X}}(F_{X_1}^{-1}(u_1), \dots, F_{X_d}^{-1}(u_d))
\end{align}

So, with a random vector $\boldsymbol{X}$ we can construct a Copula $C_{\boldsymbol{X}}$. From (\ref{copula_cdf}), if we set $u_j := F_{X_j}(x_j)$, then

\begin{equation}\label{skalar_theo}
    F(x_1, \hdots, x_d) = C(F_1(x_1), \hdots, F_{d}(x_d))
\end{equation}

This is one side of the Skalar's Theorem, which state that there exists a copula C, such that (\ref{skalar_theo}) occurs. We can derive the joint density from (\ref{skalar_theo}) by

\begin{align}\label{joint_density}
    \frac{\partial F_{\boldsymbol{X}}}{\partial \boldsymbol{x}} & = \frac{\partial C(F_{X_1}(x_1), \dots, F_{X_d}(x_d))}{\partial \boldsymbol{x}} \nonumber \\
    & = \frac{\partial C(u_1, \dots, u_n)}{\partial \boldsymbol{u}} \frac{\partial F_{X_1}}{\partial x_1} \dots \frac{\partial F_{X_d}}{\partial x_d}\hspace{0.1cm} (u_j:=F_{X_j(x_j)}) \nonumber \\
    & = c(u_1, \dots, u_n) f_{X_1}\dots f_{X_n},
\end{align}

\noindent where $f_{X_1}\dots f_{X_n}$ is the marginals densities of the random variables $X_1, \dots, X_n$, and $c(u_1, \dots, u_n)$ is the copula density given by

\begin{equation}
    c(u_1, \ldots, u_d)
    = \frac{
        f\!\left(F_1^{-1}(u_1), \ldots, F_d^{-1}(u_d)\right)
      }{
        \prod_{i=1}^d f_i\!\left(F_i^{-1}(u_i)\right)
      },
    \qquad u_i = F_i(x_i).
\end{equation}

In this work we focus on two copulas: the independence copula and Gaussian copula. The independence copula satisfies

\begin{equation}
    C(\boldsymbol{u}) = \prod_{j=1}^J u_j,
\end{equation}

\noindent and it is easy to see that the random variables $X_j's$ are independent and the joint pdf is given by

\begin{equation}
    f(x_1, \dots, x_d) = \prod_{j = 1}^d f(x_j).
\end{equation}

The Gaussian copula is defined from a multivariate Normal distribution $\Phi_{P}$ with mean vector equals to zero and correlation matrix $P$

\begin{equation}
    C_{P}^{Gauss}(u_1, \dots, u_d) = \Phi_{P}(\Phi^{-1}(u_1), \dots, \Phi^{-1}(u_d)),
\end{equation}

\noindent with density 

\begin{align}\label{gaussian_copula}
    c_{p}^{Gauss}(u_1, \dots, u_d) & = \frac{\phi_P (\Phi^{-1}(u_1), \dots, \Phi^{-1}(u_d))}{\prod_{i = 1}^d \phi(\Phi^{-1}(u_i))} \nonumber \\
    & = \frac{(2\pi)^{-n/2}|P|^{-1/2} \exp\left( -0.5 \boldsymbol{z}^{\top}(P^{-1})\boldsymbol{z}\right) }{(2\pi)^{-n/2} \exp \left(-0.5 \boldsymbol{z}^{\top}\boldsymbol{z}\right)} \nonumber \\
    & = |P|^{-1/2} \exp \left( -0.5\boldsymbol{z}^{\top}(P^{-1} - I)\boldsymbol{z}  \right)
\end{align}

\noindent with $\boldsymbol{z} = (\Phi^{-1}(u_1), \dots, \Phi^{-1}(u_d))^{\top}.$ The joint pdf of the r.v. $\boldsymbol{X}$ with arbitrary marginals and Gaussian Copula presented in (\ref{gaussian_copula}) is given by

\begin{equation}
    f(x_1, \dots, x_d) = |P|^{-1/2} \exp \left( -0.5\boldsymbol{z}^{\top}(P^{-1} - I)\boldsymbol{z}  \right) \prod_{j = 1}^d f(x_j).
\end{equation}

\section{Wavelet–Copula Variational Family}
\label{section_copula_family}

\subsection{A review on the Variational Inference}
\label{section_VI}

In this work we will consider parameterized densities for the Variational family, that is, all densities distributions have a parametric form as $q(\boldsymbol{\theta}|\boldsymbol{\psi}) \in \mathcal{Q}$, where $\mathcal{Q} = \lbrace q(\boldsymbol{\theta}|\boldsymbol{\psi}): \boldsymbol{\psi} \in \Psi\rbrace $ is a family of distributions on $\boldsymbol{\Theta}$ parameterized by $\boldsymbol{\psi}$. Then, the search for the best distribution is now for the optimal parameter in the parameter space $\Psi.$

We should look for the optimal parameter $\boldsymbol{\psi}$ which minimize the KLD between $q(\boldsymbol{\theta}|\boldsymbol{\psi})$ and the true posterior. Let $\boldsymbol{y} = (y_1, \hdots, y_n)$ for some integer $n$ and $\boldsymbol{y}$ be a real vector of some sample data, an assumed model $p(y|\boldsymbol{\theta})$ with parameters $\boldsymbol{\theta}$, $\boldsymbol{\theta} \in \boldsymbol{\Theta}$, a prior knowledge $p(\boldsymbol{\theta})$ about the vector of parameters, and the true posterior distribution of the parameters $p(\boldsymbol{\theta}|\boldsymbol{y})$, provided by the Bayes theorem. The KLD between $q_{\boldsymbol{\psi}}(\boldsymbol{\theta})$ and $p(\boldsymbol{\theta}|\boldsymbol{y})$ can be written as

\begin{align}\label{KLD}
    \text{KLD}(q(\boldsymbol{\theta}|\boldsymbol{\psi})||p(\boldsymbol{\theta}|\boldsymbol{y})) = & \mathbb{E}_{q(\boldsymbol{\theta}|\boldsymbol{\psi})}\left[\log \frac{q(\boldsymbol{\theta}|\boldsymbol{\psi})}{p(\boldsymbol{\theta}|\boldsymbol{y})}\right] = \mathbb{E}_{q(\boldsymbol{\theta}|\boldsymbol{\psi})}[\log q(\boldsymbol{\theta}|\boldsymbol{\psi})] - \mathbb{E}_{q(\boldsymbol{\theta}|\boldsymbol{\psi})}[\log p(\boldsymbol{\theta}|\boldsymbol{y})] \nonumber \\
    = &\mathbb{E}_{q(\boldsymbol{\theta}|\boldsymbol{\psi})}[\log q(\boldsymbol{\theta}|\boldsymbol{\psi})] - \mathbb{E}_{q(\boldsymbol{\theta}|\boldsymbol{\psi})}[\log p(\boldsymbol{\theta}|\boldsymbol{y})] \nonumber \\
     = &\mathbb{E}_{q(\boldsymbol{\theta}|\boldsymbol{\psi})}[\log q(\boldsymbol{\theta}|\boldsymbol{\psi})] - \mathbb{E}_{q(\boldsymbol{\theta}|\boldsymbol{\psi})}[\log p(\boldsymbol{\theta},\boldsymbol{y})] + \log p(\boldsymbol{y}) \nonumber \\
     = & - \text{ELBO}(\boldsymbol{y},\boldsymbol{\psi}) + \log p(\boldsymbol{y}).& 
\end{align}

In the expression (\ref{KLD}) $\log p(\boldsymbol{y})$ is constant, the KLD is always greater or equal than zero, so minimize (\ref{KLD}) in $\boldsymbol{\psi}$ is the same that maximize the Evidence Lower Bound (ELBO) defined as

\begin{align}
    \text{ELBO} & = \mathbb{E}_{q(\boldsymbol{\theta}|\boldsymbol{\psi})}[\log p(\boldsymbol{\theta},\boldsymbol{y})] - \mathbb{E}_{q(\boldsymbol{\theta}|\boldsymbol{\psi})}[\log q(\boldsymbol{\theta}|\boldsymbol{\psi})].
\end{align}

Therefore, the variational objective is 

\begin{equation}\label{obj}
\boldsymbol{\psi}^* = \underset{\psi}{\text{arg max ELBO}}(\boldsymbol{y}, \boldsymbol{\psi}). 
\end{equation}

The solution of (\ref{obj}), $\boldsymbol{\psi}^*$, gives the best approximate, $q(\boldsymbol{\theta}|\boldsymbol{\psi}^*)$, to the true posterior $p(\boldsymbol{\theta}|\boldsymbol{y})$ in the variational family $\mathcal{Q}$.

\subsection{Posterior densities represented by Wavelets}

The Wavelets is sufficient rich to represent a vast variety of functions, and they can produce smooth representations if it is well pruned. So, considering the notation defined in Section \ref{section_VI}, we have that our variational approximation of joint posterior density is denoted by $q(\boldsymbol{\theta}|\boldsymbol{\psi})$ for $\boldsymbol{\theta} \in \Theta$, for simplicity let us assume $\Theta = \mathbb{R}^d$, and the marginal posterior densities are $q(\theta_j|\psi_j)$ for $j = 1, \dots, d$. Consider there is a signal associated with each of these marginals that we wish to estimate, for every $j$ we have a signal of length $N = 2^J$ denoted as $\boldsymbol{q}_{j} = (q_j^{[1]}, \dots, q_j^{[N]})$ indexed by a set of integers $1, \dots, N$ with $J \in \mathbb{N}$. 

A DWT of the square root of the signal $\sqrt{\boldsymbol{q}_{j}}$ for some $j$, signal of length of $2^6 = 64$ with $J = 6$, and $R = 5$, that is, one step of filtering and down-sampling, lead us to the following coefficients $\lbrace \boldsymbol{c}_j^{(5)}, \boldsymbol{d}_j^{(5)}\rbrace$ where both vectors are length of 32. The DWT on the root square signal instead of the original signal was also used in \cite{pinheiro1997estimating}, where the authors suggest that with this transformation we do not have to worry about negative values of density

Performing a one step of IDWT in these signals leads us to the original signal $\sqrt{\boldsymbol{q}^{(j)}}$ of length 64. Instead, if we set to zero all details coefficients, only the approximating coefficients will left for the IDWT, and as presented in (\ref{IDWT}) we got

\begin{equation}\label{square_root_1}
    \sqrt{\boldsymbol{q}_{j}} \approx (H^*)\boldsymbol{c}_j^{(R)}
\end{equation}

This lead us to half of the parameters, but maintain the signal of length 64. Considering that a density functions is smooth, and in general do not have abrupt changes, we decide to represent our unknown square root signal as in (\ref{square_root_1}) using the Daubechies wavelets filters with 2 vanishing moments, more details about this family can be found in \cite{morettin2017wavelets}. 

Squaring the signal in we got $\boldsymbol{q}_j$, a new positive signal indexed in the integers. To get a sequence of evaluated density points (positive and the integral sums to 1) we associate this reconstructed signal to new axis uniformly gridded in some interval $(\delta_1, \delta_2) \in \mathbb{R}^2$, with $\delta_2 > \delta_1$. Then, the density evaluated in a grid $\lbrace \theta_j^{[1]}, \dots, \theta_{j}^{[64]} \rbrace$ point is 

\begin{equation}\label{discrete_wave_density}
    q(\theta_j^{[i]}) = \frac{(\delta_{j,2} - \delta_{j,1})\sqrt{\boldsymbol{q}_j^{[i]}}}{\sum_i\sqrt{\boldsymbol{q}_j^{[i]}}\Delta\theta_j^{[i]}}.
\end{equation}

The discrete representation of marginal variational posterior $q(\theta_j|\boldsymbol{\psi}_j)$ is (\ref{discrete_wave_density}) with the variational parameters $\boldsymbol{\psi}_j = (\delta_{j,2},\delta_{j,1}, \boldsymbol{c}_j^{(R)})$.

We propose to construct a variational family for the posterior joint distribution of vector $\boldsymbol{\theta}$ from the joint density definition in (\ref{joint_density}). From the wavelet representation of marginal densities in (\ref{discrete_wave_density}) for a set of $\theta_j$ values with parameters $\boldsymbol{c}_j, \delta_{j,1}$, and $\delta_{j,2}$, and with a copula density as defined in (\ref{joint_density}) we can write the variational posterior distribution evaluated at the grid $
\big\{ \theta_j^{[s]} \;:\; j = 1,\dots,d,\;\; s = 1,\dots,64 \big\}$
as

\begin{equation}\label{joint_wavelet_density}
q(\boldsymbol{\theta}^{[s]} \mid \boldsymbol{\zeta})
=
c(u_1, u_2, \dots, u_d \mid \boldsymbol{\xi})
\prod_{j = 1}^d
q(\theta_j^{[s]} \mid \psi_j)
\end{equation}

\noindent
where $\boldsymbol{\theta}^{[s]} \in \mathbb{R}^d$ is a a vector with components
$\theta_j^{[s]}$, $j = 1, \dots, d$.
The vector $\boldsymbol{\zeta} = (\boldsymbol{\psi}, \boldsymbol{\xi})$ collects all variational parameters,
with $\boldsymbol{\psi} = (\psi_1, \dots, \psi_d)$ denoting the parameters of the marginal distributions
and $\boldsymbol{\xi}$ the copula parameters.
Moreover, $u_j = F_j(\theta_j^{(s)})$, where $F_j(\cdot)$ is the cumulative distribution function associated with
$q(\theta_j \mid \psi_j)$.
Finally, $c(\cdot \mid \boldsymbol{\xi})$ denotes the copula density and
$q(\theta_j^{[s]} \mid \psi_j)$ is the corresponding marginal density defined in~(\ref{discrete_wave_density}) evaluated at $\theta_j^{[s]}$.

For the identity copula, (\ref{joint_wavelet_density}) reduces to
\begin{equation}
q(\boldsymbol{\theta}^{[s]} \mid \boldsymbol{\zeta})
=
\prod_{j = 1}^d
q(\theta_j^{[s]} \mid \psi_j),
\end{equation}
\noindent
with $\boldsymbol{\zeta} = \boldsymbol{\psi}$.

We sample from this variational family by simulating from all marginals using an approximate inverse transform method.
Specifically, we evaluate the marginal discrete approximation of the cumulative distribution function (CDF) associated with~$q(\theta_j^{[s]} \mid \psi_j)$ for all $j = 1, \dots, d$,
using the finite grid of points obtained from this density.
The inverse CDF is then approximated via linear interpolation.

For the Gaussian copula, (\ref{joint_wavelet_density}) becomes
\begin{equation}\label{gausian_copula_wavelet_density}
q(\boldsymbol{\theta}^{[s]} \mid \boldsymbol{\zeta})
=
|P|^{-1/2}
\exp\!\left(
-\tfrac{1}{2}
\boldsymbol{z}^{\top}
\left(P^{-1} - I\right)
\boldsymbol{z}
\right)
\prod_{j = 1}^d
q(\theta_j^{[s]} \mid \psi_j),
\end{equation}

\noindent
where
$\boldsymbol{z} =
(\Phi^{-1}(F(\theta_1^{[s]})), \dots, \Phi^{-1}(F(\theta_d^{[s]})))$,
$\Phi(\cdot)$ denotes the standard normal cumulative distribution function,
and $\boldsymbol{\zeta} = (P, \boldsymbol{\psi})$. 

The procedure to sample from (\ref{gausian_copula_wavelet_density}) is: 1) Given covariance matrix $\Sigma$, let $P$ be the corresponding correlation matrix; 2) Compute the Cholesky decomposition of $P = A^{\top}A$; 3) Generate $Z\sim N_d(\boldsymbol{0}, \boldsymbol{I}_d)$; 4) Set $\boldsymbol{Y} = A^{\top}\boldsymbol{Z}$; 5) Return $\boldsymbol{U} = (\Phi(Y_1), \dots, \Phi(Y_d))$; 6) Generate $\boldsymbol{\theta} = (F_1^{-1}(U_1), \dots, F_j^{-1}(U_d)).$, where $F_j^{-1}(\cdot)$ is the inverse CDF associated with
$q(\theta_j \mid \psi_j)$.

Then, the Evidence Lower Bound (ELBO) for this problem is defined as

\begin{align}\label{ELBO}
\mathrm{ELBO}(\boldsymbol{\zeta})
& =
\mathbb{E}_{q(\boldsymbol{\theta} \mid \boldsymbol{\zeta})}
\big[
\log p(\boldsymbol{\theta}, \boldsymbol{y})
\big]
-
\mathbb{E}_{q(\boldsymbol{\theta} \mid \boldsymbol{\zeta})}
\big[
\log q(\boldsymbol{\theta} \mid \boldsymbol{\zeta})
\big]. \nonumber  \\
& = \mathbb{E}_{q(\boldsymbol{\theta}|\boldsymbol{\zeta})}[\log p(\boldsymbol{y}|\boldsymbol{\theta}) + \log p(\boldsymbol{\theta})] - \mathbb{E}_{q(\boldsymbol{\theta}|\boldsymbol{\zeta})}\left[\log c(\boldsymbol{u}|\boldsymbol{\xi}) + \sum_j q(\theta_j|\psi_j)\right] \nonumber \\
& = \mathbb{E}_{q(\boldsymbol{\theta}|\boldsymbol{\zeta})}[\log p(\boldsymbol{y}|\boldsymbol{\theta})] + \mathbb{E}_{q(\boldsymbol{\theta}|\boldsymbol{\zeta})}[\log p(\boldsymbol{\theta})]  - \mathbb{E}_{q(\boldsymbol{\theta}|\boldsymbol{\zeta})}[\log c(\boldsymbol{u}|\boldsymbol{\xi})] -  \mathbb{E}_{q(\boldsymbol{\theta}|\boldsymbol{\zeta})}\left[\sum_j q(\theta_j|\psi_j)\right] \nonumber  \\
& = \mathbb{E}_{q(\boldsymbol{\theta}|\boldsymbol{\zeta})}[\log p(\boldsymbol{y}|\boldsymbol{\theta})] + \mathbb{E}_{q(\boldsymbol{\theta}|\boldsymbol{\zeta})}[\log p(\boldsymbol{\theta})]  - \mathbb{E}_{q(\boldsymbol{\theta}|\boldsymbol{\zeta})}[\log c(\boldsymbol{u}|\boldsymbol{\xi})] -  \sum_j \mathbb{E}_{q(\boldsymbol{\theta}|\boldsymbol{\zeta})}\left[q(\theta_j|\psi_j)\right] \nonumber   \\
& = \ell(\boldsymbol{\zeta})  - \mathbb{E}_{q(\boldsymbol{\theta}|\boldsymbol{\zeta})}[\log c(\boldsymbol{u}|\boldsymbol{\xi})] -  \sum_j \mathbb{E}_{q(\theta_j|\boldsymbol{\psi_j})}\left[q(\theta_j|\psi_j)\right] . 
\end{align}

The term $\mathbb{E}_{q(\boldsymbol{\theta}|\boldsymbol{\zeta})}[\log c(\boldsymbol{u}|\boldsymbol{\xi})]$ is zero for independent copulas and $-0.5 \log |P|$ for the Gaussian Copula. The last term in (\ref{ELBO}) is the sum of univariate integrals and can be evaluated using numerical integration. The variational parameters are estimated by solving

\begin{equation}\label{objective}
\boldsymbol{\zeta}^\star
=
\arg\max_{\boldsymbol{\zeta}}
\mathrm{ELBO}(\boldsymbol{\zeta}).
\end{equation}

The optimization problem in~(\ref{objective}) is unconstrained, since all variational parameters lie on the real domain.
For instance, we enforce the ordering constraint
$\delta_{j,2} = \delta_{j,1} + \exp(a_j)$,
which preserves the required ordering while keeping the parametrization unconstrained.
The correlation matrix $P$ is constructed from the covariance matrix $\Sigma$.
We parametrize $\Sigma$ via an unconstrained Cholesky factorization,
mapping real-valued parameters to a valid positive definite covariance matrix. So, (\ref{objective}) can be solved using a gradient-based ascent algorithm.
In a traditional setting, this would require the manual derivation of the gradient,
and in some cases the Hessian matrix, of the objective function.
This, in turn, demands analytical expressions for the gradients of the expectations appearing in~(\ref{ELBO}),
or accurate and computationally efficient approximations of the corresponding integrals.
Besides being cumbersome, such derivations are model-specific and therefore lack general applicability,
as different models lead to different gradient expressions. 

Motivated by \cite{bb_vi}, we propose a methodology that can be applied to all models considered in this work, and can be easily extend to many other models.
Specifically, we evaluate the ELBO function in (\ref{ELBO}) by drawing samples from the variational distribution, evaluating $\mathcal{L}(\boldsymbol{\zeta})$ using these samples,
and constructing Monte Carlo (MC) estimates of the corresponding expected values. The evaluated ELBO can be automatically differentiated with respect to the variational parameters
using Automatic Differentiation (AD).
AD refers to a family of techniques that compute derivatives by systematically applying the chain rule
during code execution, producing numerical evaluations of derivatives rather than symbolic expressions \cite{baydin2018automatic}. This approach requires that all transformations involving the variational parameters be differentiable.
In our setting, the parameters propagate through the following sequence of operations:
(i) inverse discrete wavelet transform (IDWT),
(ii) squaring of the reconstructed signal,
(iii) normalization,
(iv) sampling, and
(v) ELBO evaluation.
All these steps are composed of differentiable functions on the real domain.
In particular, the IDWT consists solely of linear filtering operations, as described in (\ref{IDWT}),
and therefore preserves differentiability throughout the computational graph. In this work, we implement the ELBO expression in (\ref{ELBO}) using the PyTorch library.
PyTorch is a machine learning framework that supports automatic differentiation of scalar-valued functions
\cite{paszke2017automatic}.
In addition, PyTorch provides support for the inverse discrete wavelet transform (IDWT)
through the \texttt{pytorch-wavelets} library. 

These noisy gradient estimates are then used within a stochastic optimization algorithm
to update the variational parameters in~(\ref{objective}). In the applications considered in this work, we employ the Adam optimizer \cite{adam2014method}
and RMSProp \cite{RMSProp} to optimize the variational objective. The Algorithm \ref{alg1} described the procedure using the Adam optimizer.

%
%
%
%
%
%
%
%
%
%

\begin{algorithm}[H]
\caption{Variational Inference with Wavelet Copula family}
\begin{algorithmic}[1]

\State Initialize the variational parameters
$\boldsymbol{\zeta}^{(0)} = (\boldsymbol{\xi}^{(0)}, \boldsymbol{\psi}^{(0)}, \boldsymbol{\delta}^{(0)})$.

\State Set the learning rate $\lambda$, $\boldsymbol{\beta} = (0.9,0.999), \epsilon = 10^{-8}$, $m_0 = 0, v_0 = 0$, and the maximum number of iterations
$T$ for the Adam optimizer.

\For{$t = 1, \dots, T$}

  \State Construct a uniform grid
  $\boldsymbol{\theta}_j^* = \{\theta_j^{[1]}, \dots, \theta_j^{[64]}\}$
  over the interval $(\delta_{j,1}^{(t)}, \delta_{j,2}^{(t)})$,
  for $j = 1, \dots, d$.

  \State Evaluate the marginal densities
  $q(\theta_j \mid \boldsymbol{\psi}^{(t)})$
  on the grid $\boldsymbol{\theta}_j^*$.

  \State Compute the marginal CDFs
  $F_j(\theta_j \mid \boldsymbol{\psi}^{(t)})$ on the grid $\boldsymbol{\theta}_j^*$.

  \State Draw $S$ samples
  $\boldsymbol{\theta}^{(1)}, \dots, \boldsymbol{\theta}^{(S)}$
  from the joint variational density
  $q(\boldsymbol{\theta} \mid \boldsymbol{\zeta}^{(t)})$.

  \State Estimate the expected log joint density

\[
\hat{\ell}^{(t)}
=
\frac{1}{S}
\sum_{s=1}^S
\log p\bigl(\boldsymbol{y}, \boldsymbol{\theta}^{(s)}\bigr),
\qquad
\boldsymbol{\theta}^{(s)} \sim q\bigl(\boldsymbol{\theta}\mid\boldsymbol{\zeta}^{(t)}\bigr).
\]

\State Compute $\mathbb{E}_{q(\boldsymbol{\theta}|\boldsymbol{\zeta})}[\log c(\boldsymbol{u}|\boldsymbol{\xi})]$ and $\sum_j \left[\sum_iq(\theta_j^{[i]}|\psi_j)\right]\Delta \boldsymbol{\theta}_j^* $

  \State Compute the gradient

\Statex
\[
g^{(t)}
=
\frac{1}{S}
\sum_{s=1}^S
\nabla_{\boldsymbol{\zeta}}
\log p\bigl(\boldsymbol{y}, \boldsymbol{\theta}^{(s)}\bigr) - \nabla_{\zeta} \mathbb{E}_{q(\boldsymbol{\theta}|\boldsymbol{\zeta})}[\log c(\boldsymbol{u}|\boldsymbol{\xi})] -
\sum_j\nabla_{\boldsymbol{\psi_j}} \left[\sum_iq(\theta_j^{[i]}|\psi_j)\right]\Delta \boldsymbol{\theta}_j^*.
\]

\State Update $\boldsymbol{\zeta}^{(t)}$ using the Adam recursion:
\Statex
\[
\begin{aligned}
m_t &= \beta_1 m_{t-1} + (1 - \beta_1)\, g^{(t)}, \\
v_t &= \beta_2 v_{t-1} + (1 - \beta_2)\, \bigl(g^{(t)}\bigr)^2, \\
\hat{m}_t &= \frac{m_t}{1 - \beta_1^t}, \qquad
\hat{v}_t = \frac{v_t}{1 - \beta_2^t}, \\
\boldsymbol{\zeta}^{(t)} &=
\boldsymbol{\zeta}^{(t - 1)}
- \lambda \frac{\hat{m}_t}{\sqrt{\hat{v}_t} + \epsilon}.
\end{aligned}
\]

\State \Return $\boldsymbol{\zeta}^{(T)}$

\EndFor

\end{algorithmic}
\label{alg1}
\end{algorithm}

\section{Simulation study}

\subsection{Simulation study 1: Logistic Regression}

In this first simulation study, we consider a generalized linear model with a binary response. Let $\boldsymbol{y} = (y_1,\ldots,y_n)$ denote the vector of binary outcomes and $\boldsymbol{x}_i$ the corresponding vector of covariates for observation $i$. The goal of this simulation is to assess the performance of the proposed methodology in recovering the regression coefficients under a standard logistic regression setting.

We assume a Bernoulli likelihood with a logit link function and place a weakly informative Gaussian prior on the regression coefficients $\boldsymbol{\beta}$. The hierarchical model is specified as

\begin{align}\label{model1_simu}
    y_i \mid \boldsymbol{\beta} &\sim \text{Bernoulli}\!\left(\text{logit}^{-1}(\boldsymbol{x}_i^\top \boldsymbol{\beta})\right), \nonumber \\
    \boldsymbol{\beta} &\sim \mathcal{N}\!\left(\boldsymbol{0},\, 100\,\boldsymbol{I}\right).
\end{align}

\noindent for $i = 1, \hdots, p$, and $\boldsymbol{\beta}$ of length $p$. We considered six simulation scenarios defined by different combinations of the number of covariates $p$ and the sample size $n$: $p=10$ with $n=200$ and $n=1000$; $p=20$ with $n=1000$ and $n=5000$; and $p=50$ with $n=5000$. For each scenario, we generated $R=100$ independent replications of the model. We benchmarked the proposed Wavelet–Copula variational family, considering both
independent and Gaussian copulas, against the No-U-Turn Sampler (NUTS)
implemented in Stan, using CmdStanPy as the interface. The variational procedure
follows Algorithm~\ref{alg1}, with a learning rate of 0.01 and 50 Monte Carlo
samples per iteration.

We evaluated the performance of the competing methods by summarizing, across replications, the posterior mean and posterior standard deviation of each regression coefficient. In addition, we assessed the mean absolute error (MAE) for each coefficient $\beta_i$, $i=1,\ldots,p$, defined as
$
\mathrm{MAE}(\hat{\beta}_i)
=
\frac{1}{R}
\sum_{r=1}^{R}
\lvert \hat{\beta}_{i,r} - \beta_i \rvert,$
where $\hat{\beta}_{i,r}$ denotes the posterior mean of the $i$th coefficient obtained from the $r$th replication.

Table \ref{table1_simu1_n1000} reports the performance metrics for the scenarios with $p=10$ and sample sizes $n=200$ and $n=1000$, considering the independent copula. Posterior mean estimates obtained via VI closely match those from the MCMC approach implemented in Stan across all parameters. A similar agreement is observed for the posterior standard deviations; however, the variational posterior slightly underestimates uncertainty relative to MCMC. The MAE values are comparable across both approaches, indicating that the proposed wavelet-based variational family achieves estimation accuracy similar to that of MCMC posterior mean estimates in these scenarios.

\begin{table}
\centering
\caption{True values ($\boldsymbol{\beta}$), average posterior mean (Mean) and posterior standard deviation (SD) obtained from variational Bayes with the Wavelet-copula variational family with independent copula and from NUTS algorithm in Stan, as well as the MAE for all regression coefficients of model (\ref{model1_simu}).}
\begin{tabular}{r|ccc|ccc}
\hline
& \multicolumn{6}{c}{$p = 10, n = 200$} \\
\hline
& \multicolumn{3}{|c|}{VB + Wavelets} & \multicolumn{3}{c}{NUTS} \\ 
 True values & Mean & SD & MAE & Mean & SD & MAE \\
\hline
$(\beta_0)$ -0.111 & -0.174 & 0.190 & 0.172 & -0.176 & 0.197 & 0.170 \\
$(\beta_1)$ 0.120 & 0.127 & 0.193 & 0.163 & 0.128 & 0.199 & 0.165 \\
$(\beta_2)$ -0.370 & -0.441 & 0.192 & 0.178 & -0.437 & 0.203 & 0.174 \\
$(\beta_3)$ -0.240 & -0.262 & 0.192 & 0.164 & -0.260 & 0.202 & 0.161 \\
$(\beta_4)$ -1.197 & -1.415 & 0.217 & 0.284 & -1.404 & 0.253 & 0.278 \\
\hline
& \multicolumn{6}{c}{$p = 10, n = 1000$} \\
\hline
& \multicolumn{3}{|c|}{VB + Wavelets} & \multicolumn{3}{c}{NUTS} \\ 
 True values & Mean & SD & MAE & Mean & SD & MAE \\
\hline
-0.111 $(\beta_0)$ & -0.110 & 0.079 & 0.066 & -0.111 & 0.081 & 0.066 \\
0.120  $(\beta_1)$ & 0.133 & 0.079 & 0.067 & 0.133 & 0.081 & 0.067 \\
-0.370 $(\beta_2)$ & -0.377 & 0.081 & 0.058 & -0.376 & 0.083 & 0.058 \\
-0.240 $(\beta_3)$ & -0.243 & 0.080 & 0.069 & -0.244 & 0.082 & 0.069 \\
-1.197 $(\beta_4)$ & -1.247 & 0.092 & 0.095 & -1.250 & 0.101 & 0.095 \\
\hline
\end{tabular}
\label{table1_simu1_n1000}
\end{table}

\begin{figure}
    \centering
    \includegraphics[width=0.7\linewidth]{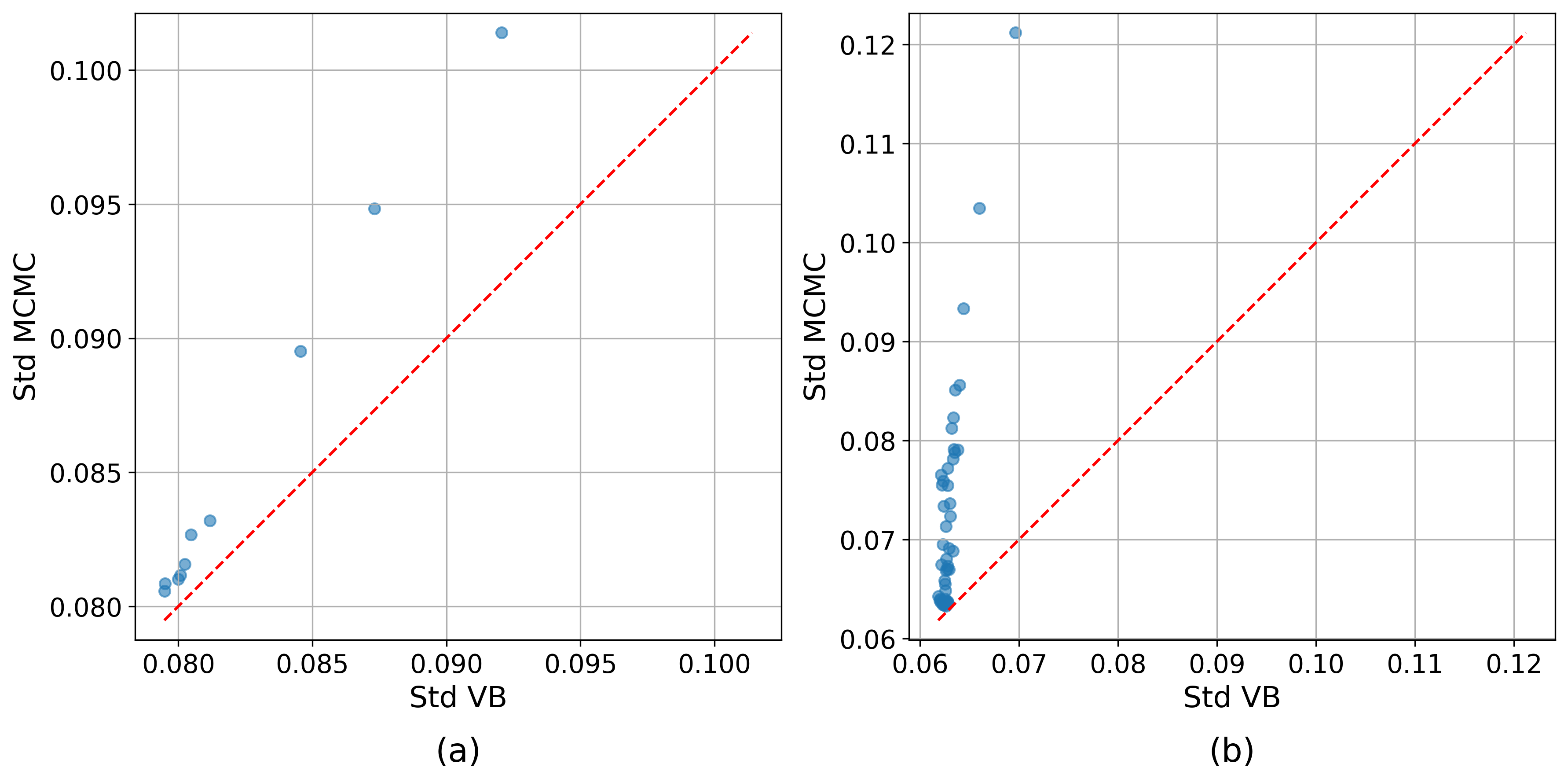}
    \caption{Posterior standard deviation (Std) from MCMC method (y) and VB method with mean-field Wavelet family (x) for the scenario with $n = 1000$ and $p = 10$ (label (a)) and with $n = 5000$ and $p = 50$ (label (b)).}
    \label{sd_meanfield_wavelets1}
\end{figure}

Tables~\ref{table1_simu1_p20} and~\ref{table1_simu1_p50} report the performance metrics for the remaining scenarios under the independent copula. The conclusions regarding posterior mean estimation and MAE are consistent with those reported in Table~\ref{table1_simu1_n1000}, indicating that the methodology adapts well to larger sample sizes and an increasing number of covariates. However, the independent copula family continues to underestimate posterior variability. Figures~\ref{sd_meanfield_wavelets1}(a) and~(b) display the posterior standard deviations obtained from MCMC and VI under the independent copula family. A clear underestimation of posterior variability is observed, particularly in the scenario with $p=50$ and $n=5000$.

\begin{table}
\centering
\caption{True values ($\boldsymbol{\beta}$), average posterior mean (Mean) and posterior standard deviation (SD) obtained from variational Bayes with the Wavelet-copula variational family with independent copula and from NUTS algorithm in Stan, as well as the MAE for the first 10 regression coefficients of model (\ref{model1_simu}).}
\begin{tabular}{r|ccc|ccc}
\hline
& \multicolumn{6}{c}{$p = 20, n = 1000$} \\
\hline
& \multicolumn{3}{|c|}{VB + Wavelets} & \multicolumn{3}{c}{NUTS} \\ 
 True values & Mean & SD & MAE & Mean & SD & MAE \\
\hline
0.337  ($\beta_0$)& 0.380 & 0.119 & 0.106 & 0.379 & 0.124 & 0.106 \\
-0.178 ($\beta_1$)& -0.210 & 0.120 & 0.117 & -0.211 & 0.124 & 0.117 \\
-0.304 ($\beta_2$)& -0.311 & 0.119 & 0.109 & -0.311 & 0.124 & 0.109 \\
-0.588 ($\beta_3$)& -0.650 & 0.120 & 0.126 & -0.650 & 0.130 & 0.126 \\
1.581  ($\beta_4$)& 1.735 & 0.125 & 0.189 & 1.730 & 0.168 & 0.189 \\
1.301  ($\beta_5$)& 1.448 & 0.123 & 0.175 & 1.444 & 0.155 & 0.175 \\
1.275  ($\beta_6$)& 1.411 & 0.124 & 0.165 & 1.408 & 0.154 & 0.165 \\
-0.201 ($\beta_7$)& -0.233 & 0.119 & 0.113 & -0.234 & 0.123 & 0.113 \\
-0.161 ($\beta_8$)& -0.157 & 0.118 & 0.096 & -0.156 & 0.122 & 0.096 \\
-0.401 ($\beta_9$)& -0.450 & 0.120 & 0.111 & -0.446 & 0.126 & 0.111 \\
\hline
& \multicolumn{6}{c}{$p = 20, n = 5000$} \\
\hline
& \multicolumn{3}{|c|}{VB + Wavelets} & \multicolumn{3}{c}{NUTS} \\ 
 True values & Mean & SD & MAE & Mean & SD & MAE \\
\hline
0.337  ($\beta_0$) & 0.345 & 0.052 & 0.041 & 0.344 & 0.053 & 0.041 \\
-0.178 ($\beta_1$) & -0.175 & 0.051 & 0.041 & -0.174 & 0.052 & 0.041 \\
-0.304 ($\beta_2$) & -0.299 & 0.052 & 0.043 & -0.299 & 0.052 & 0.043 \\
-0.588 ($\beta_3$) & -0.608 & 0.052 & 0.049 & -0.606 & 0.055 & 0.049 \\
1.581  ($\beta_4$) & 1.612 & 0.054 & 0.068 & 1.607 & 0.070 & 0.068 \\
1.301  ($\beta_5$) & 1.324 & 0.053 & 0.052 & 1.320 & 0.065 & 0.052 \\
1.275  ($\beta_6$) & 1.308 & 0.053 & 0.064 & 1.303 & 0.064 & 0.064 \\
-0.201 ($\beta_7$) & -0.202 & 0.052 & 0.047 & -0.201 & 0.052 & 0.047 \\
-0.161 ($\beta_8$) & -0.161 & 0.051 & 0.039 & -0.161 & 0.052 & 0.039 \\
-0.401 ($\beta_9$) & -0.419 & 0.052 & 0.045 & -0.417 & 0.053 & 0.045 \\
\hline
\end{tabular}
\label{table1_simu1_p20}
\end{table}

\begin{table}
\centering
\caption{True values ($\boldsymbol{\beta}$), average posterior mean (Mean) and posterior standard deviation (SD) obtained from variational Bayes with the Wavelet-copula variational family with independent copula and from NUTS algorithm in Stan, as well as the MAE for the first 10 regression coefficients of model (\ref{model1_simu}).}
\begin{tabular}{r|ccc|ccc}
\hline
& \multicolumn{6}{c}{$p = 50, n = 5000$} \\
\hline
& \multicolumn{3}{|c|}{VB + Wavelets} & \multicolumn{3}{c}{NUTS} \\ 
 True values & Mean & SD & MAE & Mean & SD & MAE \\
\hline
0.337  ($\beta_0$)& 0.358 & 0.062 & 0.049 & 0.356 & 0.064 & 0.049 \\
-0.178 ($\beta_1$)& -0.188 & 0.062 & 0.049 & -0.188 & 0.064 & 0.049 \\
-0.304 ($\beta_2$)& -0.325 & 0.063 & 0.060 & -0.322 & 0.064 & 0.060 \\
-0.588 ($\beta_3$)& -0.633 & 0.063 & 0.073 & -0.628 & 0.067 & 0.073 \\
0.349  ($\beta_4$)& 0.362 & 0.062 & 0.052 & 0.360 & 0.064 & 0.052 \\
0.660  ($\beta_5$)& 0.703 & 0.063 & 0.065 & 0.700 & 0.067 & 0.065 \\
-0.220 ($\beta_6$)& -0.230 & 0.063 & 0.053 & -0.230 & 0.064 & 0.053 \\
-0.379 ($\beta_7$)& -0.397 & 0.063 & 0.061 & -0.395 & 0.065 & 0.061 \\
0.767  ($\beta_8$)& 0.811 & 0.063 & 0.070 & 0.807 & 0.069 & 0.070 \\
-1.193 ($\beta_9$)& -1.273 & 0.062 & 0.096 & -1.265 & 0.077 & 0.096 \\
\hline
\end{tabular}
\label{table1_simu1_p50}
\end{table}

\begin{figure}
    \centering
    \includegraphics[width=0.7\linewidth]{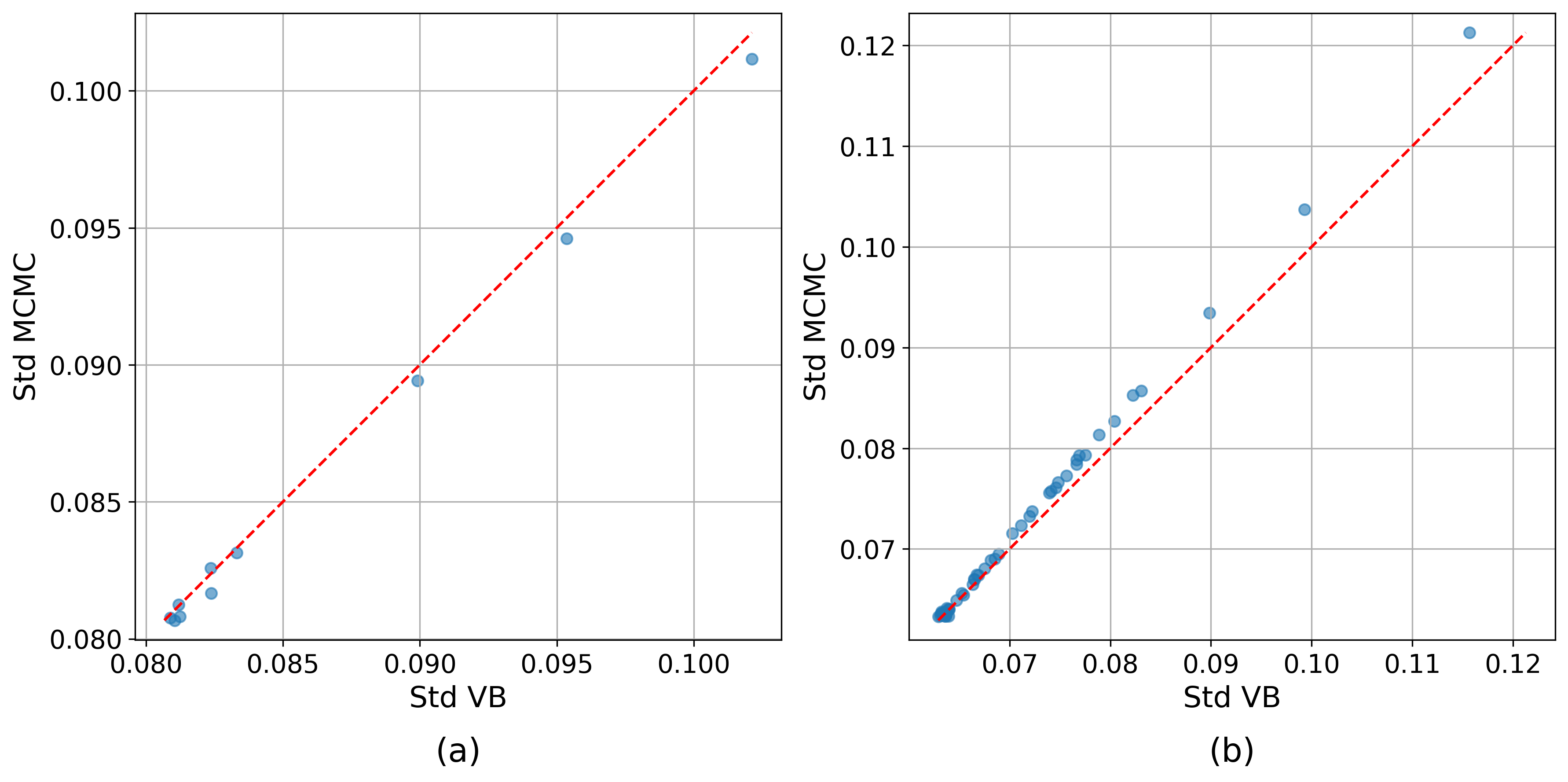}
    \caption{Posterior standard deviation (Std) from MCMC method (y) and VB method with Wavelet copula Gaussian family (x) for the scenario with $n = 1000$ and $p = 10$ (label (a)) and with $n = 5000$ and $p = 50$ (label (b)).}
    \label{sd_gaussian_wavelets1}
\end{figure}

To address the underestimation of posterior variance, we employed the Wavelet-copula family with a Gaussian copula dependence structure. Posterior mean estimates and MAE values remain comparable to those obtained via MCMC, as reported in Table~\ref{table1_simu1_p50_gaussian}, which presents results for the scenario with $n=5000$ and $p=50$. Figure~\ref{sd_gaussian_wavelets1} shows that the posterior standard deviations obtained under the Gaussian-copula family are substantially closer to the corresponding MCMC estimates.

\begin{table}[]
    \centering
    \caption{True values ($\beta$), posterior mean and standard deviation from VB with Wavelet copula Gaussian family and MCMC with Stan, and RMSE for all 
$\beta$ parameters of the model (\ref{model1_simu}) with $p = 50$, and $n = 5000$.}
    \begin{tabular}{l|rrrrrr}
\hline
& \multicolumn{6}{c}{$p = 50, n = 5000$} \\
\hline
& \multicolumn{3}{|c|}{VB + Wavelets} & \multicolumn{3}{c}{NUTS} \\ 
 True values & Mean & SD & MAE & Mean & SD & MAE \\
\hline
0.337  ($\beta_0$)& 0.355 & 0.064 & 0.049 & 0.355 & 0.064 & 0.049 \\
-0.178 ($\beta_1$)& -0.189 & 0.064 & 0.047 & -0.189 & 0.063 & 0.047 \\
-0.304 ($\beta_2$)& -0.324 & 0.064 & 0.058 & -0.323 & 0.064 & 0.058 \\
-0.588 ($\beta_3$)& -0.629 & 0.066 & 0.070 & -0.628 & 0.066 & 0.070 \\
0.349  ($\beta_4$)& 0.360 & 0.064 & 0.050 & 0.360 & 0.064 & 0.050 \\
0.660  ($\beta_5$)& 0.698 & 0.067 & 0.063 & 0.697 & 0.067 & 0.063 \\
-0.220 ($\beta_6$)& -0.231 & 0.064 & 0.055 & -0.230 & 0.064 & 0.055 \\
-0.379 ($\beta_7$)& -0.394 & 0.065 & 0.061 & -0.394 & 0.065 & 0.061 \\
0.767  ($\beta_8$)& 0.806 & 0.068 & 0.067 & 0.807 & 0.069 & 0.067 \\
-1.193 ($\beta_9$)& -1.266 & 0.075 & 0.090 & -1.266 & 0.077 & 0.090 \\
\hline
\end{tabular}
    \label{table1_simu1_p50_gaussian}
\end{table}

\subsection{Simulation study 2: Linear regression with automatic automatic relevance determination }

This simulation study considers a linear regression model with hierarchical prior distributions that induce sparsity in the regression coefficients via automatic relevance determination \cite{drugowitsch2013variational}. The hierarchical structure of the model is given by

\begin{align}\label{model2_simu}
    y_i \mid \boldsymbol{\beta}, \sigma^2 
    &\sim \mathcal{N}\!\left(\boldsymbol{x}_i^\top \boldsymbol{\beta}, \sigma^2\right), \nonumber \\
    \boldsymbol{\beta} \mid \boldsymbol{\alpha}, \sigma^2
    &\sim \mathcal{N}_P\!\left(\boldsymbol{0}, \sigma^2 \, \mathrm{diag}(\boldsymbol{\alpha})^{-1}\right), \\
    \alpha_j 
    &\sim \mathrm{Gamma}(a,b), \quad j = 1,\ldots,p, \nonumber \\
    \sigma^2 
    &\sim \mathrm{Inv\text{-}Gamma}(c,d). \nonumber
\end{align}

\noindent for $n = 1, \hdots, N,$ $j = 1, \hdots, p$, $a = b = 0.001$, and $c = d = 0.01$. We considered 18 different scenarios, sample size $n = 500$ with $p = 250$ and $500$, $n = 1000$ with $P = 250, 500$ and $1000$, and $n = 2000$ with $p = 250, 500$ and $1000$, and all these previous combinations with two levels of sparsity, $r = 0.5$ and $0.8$. The sparsity level $r$ determines the proportion of regressors with no predictive power. For example, when $r = 0.5$, half of the regressors have coefficients equal to zero and therefore do not contribute to the response. We simulate the $\boldsymbol{\beta}$ coefficients from a normal random variable with mean zero and standard deviation equals to one, and set $r \times p$ of them to zero randomly.

We bench-marked our proposal methodology, considering the independent copula for the variational family against the Automatic Differentiation Variational Inference (ADVI) implemented in Stan \cite{carpenter2017stan}. For this model, we employ the L-BFGS optimizer. This choice is motivated by the fact that the integrals involved in (\ref{ELBO}) admit accurate numerical approximations due to the independence structure of the variational posterior, together with the Gaussian form of both the likelihood and the prior.

\begin{figure}
    \centering
    \includegraphics[width=0.95\linewidth]{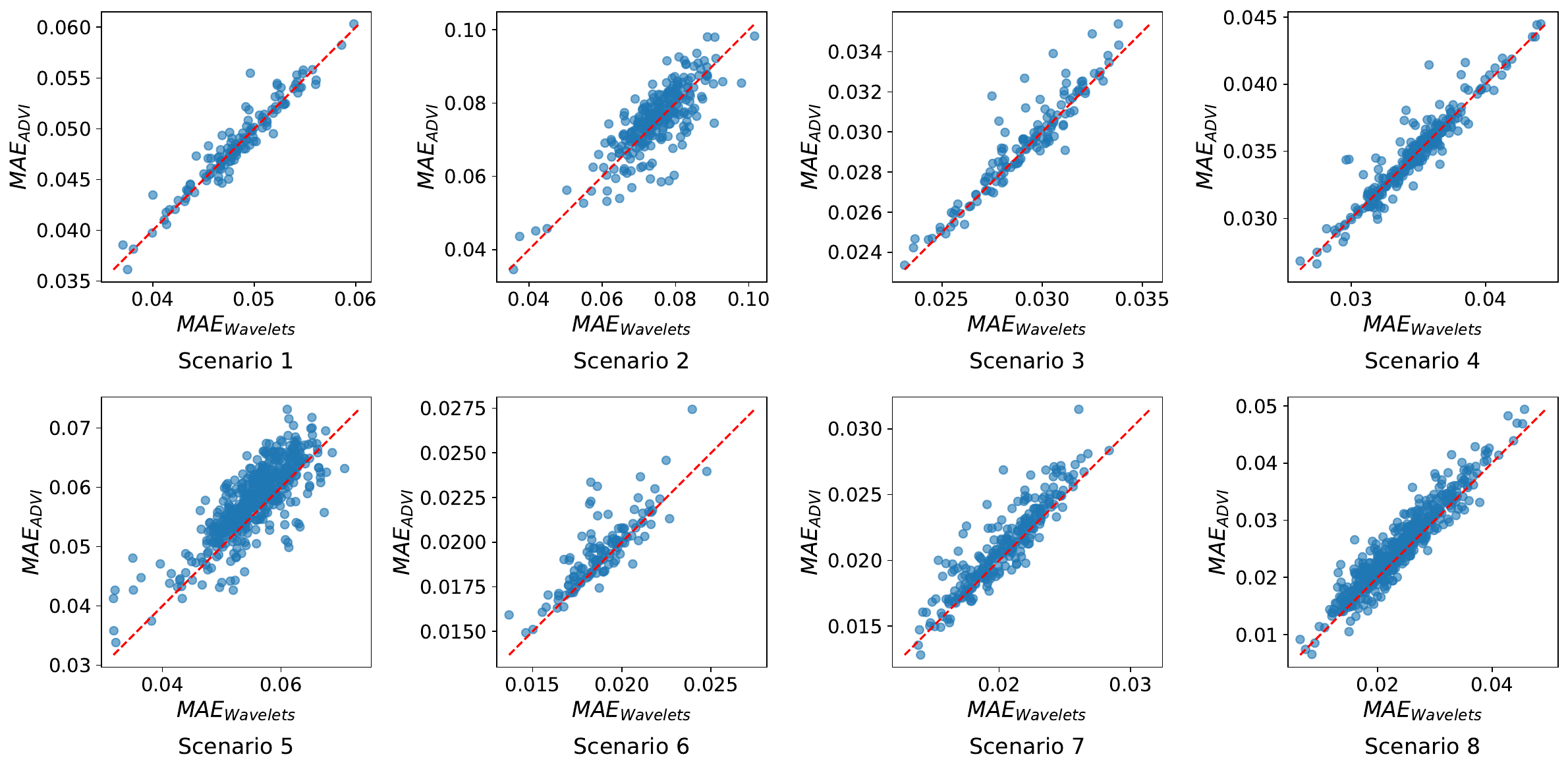}
    \caption{Mean Absolute Error (MAE) of the model parameters across scenarios, (a), for Wavelet family and ADVI.}
    \label{MAE_05}
\end{figure}

Figure~\ref{MAE_05} displays eight scatter plots, each corresponding to a different simulation scenario. In each panel, every point represents the mean absolute error (MAE) of a single regression coefficient, with the MAE obtained under the wavelet-copula variational family shown on the $x$-axis and the MAE obtained under ADVI shown on the $y$-axis. The red diagonal line corresponds to equality of MAE between the two methods. Points lying above the diagonal indicate smaller MAE for the wavelet-copula variational family, whereas points below the diagonal indicate smaller MAE for ADVI. Overall, the figure shows that the proposed variational family with an independent copula yields competitive—and in many scenarios lower—MAE values for the relevant parameters.

\begin{figure}
    \centering
    \includegraphics[width=0.8\linewidth]{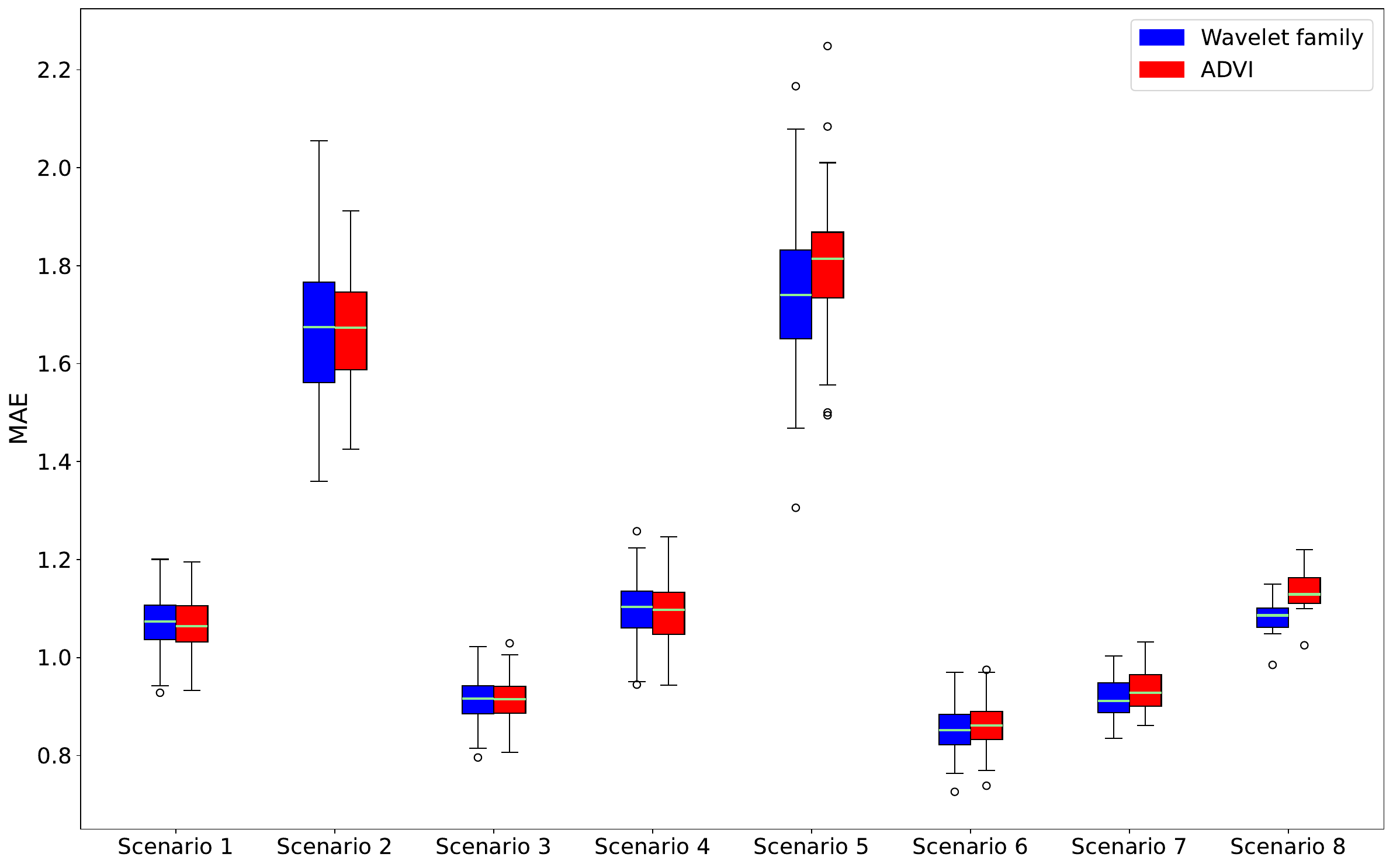}
    \caption{Mean Absolute Error (MAE) in a test set across scenarios, from one to eight, for Wavelet family and ADVI.}
    \label{MAE_test_05}
\end{figure}

In addition, we evaluated predictive performance on a test set $\boldsymbol{y}^{\text{test}}$ using the mean absolute error (MAE), defined as
\[
\mathrm{MAE}_r
=
\frac{1}{N_{\text{test}}}
\sum_{l=1}^{N_{\text{test}}}
\lvert \tilde{y}_{l,r} - y^{\text{test}}_{l} \rvert,
\]
where $\tilde{y}_{l,r}$ denotes the posterior predictive estimate for the $l-$th observation in the test set obtained from the $r$th replication. Figure~\ref{MAE_test_05} presents boxplots of the MAE across replications for all scenarios. Across most scenarios, the competing methods yield nearly identical median MAE values. In particular, in scenarios 5 and 8, the proposed wavelet-copula variational family achieves improved predictive accuracy relative to ADVI, although with increased variability in scenario 5.

The model in~\eqref{model2_simu} induces shrinkage on noninformative coefficients through the penalty parameters $\alpha_j$. Larger values of $\alpha_j$ correspond to smaller prior variances, thereby shrinking the associated regression coefficients toward zero. To identify relevant predictors, we examine the posterior means of the $\alpha$ coefficients. Let $\hat{\boldsymbol{\alpha}} = (\hat{\alpha}_1,\ldots,\hat{\alpha}_p)$ denote the vector of posterior mean estimates, and let $q_{0.5}(\boldsymbol{\alpha})$ denote its empirical median. We define the set of selected predictors as
\[
\mathcal{K}
=
\left\{ k : \hat{\alpha}_k > q_{0.5}(\boldsymbol{\alpha}),\; 1 \le k \le p \right\}.
\]
The set $\mathcal{K}$ therefore indexes the predictors selected by the model. This threshold provides a data-driven and scale-invariant criterion for variable selection.

Table~\ref{tab:placeholder} summarizes the variable selection performance across scenarios with $r=0.5$. Overall, the wavelet-copula variational family achieves a favorable balance between true positives and false discoveries, consistently yielding lower numbers of false positives and false negatives compared to ADVI across most scenarios. This behavior results in higher or comparable numbers of true negatives, indicating more parsimonious variable selection.

In moderate-dimensional settings (Scenarios~1--4 and~6--7), the wavelet-based approach exhibits slightly improved selection accuracy relative to ADVI, with fewer misclassified predictors. In high-dimensional scenarios ($p=1000$), both methods show increased difficulty in variable selection; however, while ADVI attains a larger number of true positives in Scenario~5, this improvement is accompanied by a substantial increase in false positives, suggesting over-selection. In contrast, the proposed variational family maintains a more balanced trade-off between sensitivity and specificity across all scenarios.

\begin{table}[]
    \centering
\caption{Variable selection performance for scenarios with $r=0.5$. TP, FP, FN, and TN correspond to the numbers of true positives, false positives, false negatives, and true negatives, respectively, averaged across replications.}
\begin{tabular}{l|rrrr|rrrr}
\hline
& \multicolumn{4}{c|}{Wavelet family} & \multicolumn{4}{c}{ADVI} \\
& TP & FP & FN & TN  & TP & FP & FN & TN \\ 
\hline
\makecell[l]{Scenario 1\\$p = 250,\ r = 0.5$}  & 117.286 & 7.714 & 7.714 & 116.286 & 117.780 & 7.220 & 8.220 & 116.780 \\
\hline
\makecell[l]{Scenario 2\\$p = 500,\ r = 0.5$}  & 220.767 & 29.233 & 29.233 & 219.767 & 218.744 & 31.256 & 32.256 & 217.744 \\
\hline
\makecell[l]{Scenario 3\\$p = 250,\ r = 0.5$}  & 119.303 & 5.697 & 5.697 & 118.303 & 120.112 & 4.888 & 5.888 & 119.112 \\
\hline
\makecell[l]{Scenario 4\\$p = 500,\ r = 0.5$}  & 233.357 & 16.643 & 16.643 & 232.357 & 232.107 & 17.893 & 18.893 & 231.107 \\
\hline
\makecell[l]{Scenario 5\\$p = 1000,\ r = 0.5$} & 413.943 & 86.057 & 85.057 & 413.943 & 454.182 & 45.818 & 45.818 & 454.182 \\
\hline
\makecell[l]{Scenario 6\\$p = 250,\ r = 0.5$}  & 118.224 & 6.776 & 5.776 & 118.224 & 115.463 & 9.537 & 9.537 & 115.463 \\
\hline
\makecell[l]{Scenario 7\\$p = 500,\ r = 0.5$}  & 239.926 & 10.074 & 10.074 & 238.926 & 235.667 & 14.333 & 15.333 & 234.667 \\
\hline
\makecell[l]{Scenario 8\\$p = 1000,\ r = 0.5$} & 472.143 & 27.857 & 26.857 & 472.143 & 466.143 & 33.857 & 33.857 & 466.143 \\
\hline
\end{tabular}
    \label{tab:placeholder}
\end{table}

\subsection{Simulation study 3: Hierarchical regression}

In this example, we again consider a linear regression model, now equipped with a hierarchical prior that induces correlation between random effects. Specifically, we consider a model with two random effects $\boldsymbol{b}$, whose covariance structure is governed by a matrix $\Sigma$. The hierarchical model is given by
\begin{align}
y_{ij} \mid \boldsymbol{\beta}, \boldsymbol{b}, \sigma^2 
&\sim \mathcal{N}\!\left(\boldsymbol{x}_i^\top \boldsymbol{\beta} + \boldsymbol{z}_i^\top \boldsymbol{b},\, \sigma^2 \right), \nonumber \\
\boldsymbol{b} \mid \Sigma 
&\sim \mathcal{N}_2(\boldsymbol{0},\, \Sigma), \\
\boldsymbol{\beta} 
&\sim \mathcal{N}_P(\boldsymbol{0},\, 100\, I_P). \nonumber
\end{align}

The random effects vector is defined as $\boldsymbol{b} = (b_{1}, b_{2})^{\top}$, where $(b_{1i}, b_{2i})$ are correlated. The covariance matrix $\Sigma$ is parameterized as $\Sigma = D R D$, with $D = \mathrm{diag}(\sigma_1, \sigma_2)$ and $R$ a $2\times2$ correlation matrix with correlation parameter $\rho \in (-1,1)$. This decomposition separates marginal variances from the dependence structure, facilitating interpretation and prior specification.

Gamma priors with shape and scale parameters equal to $0.01$ are assigned to $\sigma$, $\sigma_1$, and $\sigma_2$, reflecting weakly informative prior beliefs on variance components. A uniform prior is assumed for the correlation parameter $\rho$.

We simulate datasets under this model considering 12 distinct scenarios, defined by combinations of the number of individuals $n_{\text{ind}} \in \{100, 200\}$, the number of repeated measurements per individual $n_{\text{rep}} \in \{10, 20\}$, and the correlation between random effects $\rho \in \{-0.7, 0, 0.7\}$. These values correspond to covariance terms $\sigma_{12} = \rho \sigma_1 \sigma_2 \in \{-0.35, 0, 0.35\}$. The regression coefficients are fixed at $\boldsymbol{\beta} = (-2.0, 1.5)^{\top}$, with $\sigma = 1$, $\sigma_1 = 1$, and $\sigma_2 = 0.25$.

These design choices aim to assess the performance of the proposed variational family under increasing data availability and varying dependence structures. The values of $n_{\text{ind}}$ and $n_{\text{rep}}$ allow us to separately evaluate the effect of the number of subjects and the number of repeated measurements on posterior inference. The correlation levels $\rho \in \{-0.7, 0, 0.7\}$ represent strong negative, null, and strong positive dependence between random effects, respectively, providing a challenging test bed for approximations that must capture posterior dependence. 

In Table \ref{resultados_rho0_sim3}, we display the average posterior mean and posterior standard deviation, and the MAE of the estimated posterior mean of the fixed parameters and hyperparameters for all considered scenarios with $ \rho = 0$. We can see that for the fixed parameters, $\beta_0$ and $\beta_1$ the average and MAE are similar for all values of individuals an repetition. Besides that we observe a smaller MAE when we increase the number of repetition, as expected. For the hyperparameters we observe that our Wavelet-copula family produces similar results to the NUTS, indeed, smaller MAE for the scenarios with less repetitions. On the other hand, the ADVI got higher MAE for the hyperparamters in all scenarios. The average posterior standard deviation indicates that our approach slight underestimate the variance for these particularly scenarios.
Table \ref{resultados_rho07_sim3} present similar conclusion, highighting again the good results presented in our approache for the hyperparamters.

\begin{table}
\centering
    \caption{True values of the parameters, average posterior mean (Mean) and average standard deviation (SD), and MAE for all parameters of the model (\ref{model1_simu}) in different scenarios with $\rho = 0$.}
\begin{tabular}{r|rrr|rrr|rrr}
\hline
& \multicolumn{9}{c}{$n_{ind} = 100, n_{rep} = 10$} \\
\hline
& \multicolumn{3}{c|}{VB + Wavelets} & \multicolumn{3}{c|}{NUTS} & \multicolumn{3}{c}{ADVI} \\ 
 True values & Mean & SD & MAE & Mean & SD & MAE & Mean & SD & MAE \\
\hline
 -2.000 ($\beta_0$) & -1.992 & 0.083 & 0.082 & -1.993 & 0.107 & 0.084 & -1.960 & 0.122 & 0.111 \\ 
 1.500 ($\beta_1$) & 1.493 & 0.057 & 0.047 & 1.494 & 0.061 & 0.046   & 1.491 & 0.065 & 0.048 \\ 
 1.000 ($\sigma_e$) & 1.003 & 0.023 & 0.020 & 1.003 & 0.025 & 0.021 & 1.042 & 0.032 & 0.046\\
 1.000 ($\sigma_1$) & 1.030 & 0.150 & 0.131 & 1.052 & 0.171 & 0.136 & 0.844 & 0.118 & 0.169\\
 0.250 ($\sigma_2$) & 0.250 & 0.037 & 0.039 & 0.256 & 0.056 & 0.040 0 & 0.195 & 0.033 & 0.059 \\
\hline
& \multicolumn{9}{c}{$n_{ind} = 100, n_{rep} = 20$} \\
\hline
-2.000 ($\beta_0$)  & -1.978 & 0.073 & 0.099 & -1.997 & 0.104 & 0.081 & -1.974 & 0.107 & 0.095 \\ 
1.500 ($\beta_1$)  & 1.496 & 0.050 & 0.050 & 1.500 & 0.055 & 0.045   & 1.498 & 0.057 & 0.045 \\
 1.000 ($\sigma_e$) & 1.015 & 0.018 & 0.027 & 1.002 & 0.017 & 0.013 & 1.020 & 0.020 & 0.025 \\
1.000 ($\sigma_1$) & 1.023 & 0.151 & 0.121 & 1.051 & 0.161 & 0.123 & 0.816 & 0.095 & 0.189 \\
0.250 ($\sigma_2$) & 0.253 & 0.039 & 0.036 & 0.252 & 0.045 & 0.031 & 0.202 & 0.026 & 0.050 \\
\hline
& \multicolumn{9}{c}{$n_{ind} = 200, n_{rep} = 10$} \\
\hline
 -2.000 ($\beta_0$)  & -2.002 & 0.050 & 0.061 & -2.002 & 0.075 & 0.058 & -1.972 & 0.087 & 0.084 \\ 
 1.500 ($\beta_1$)  & 1.498 & 0.039 & 0.038 & 1.498 & 0.043 & 0.038   & 1.495 & 0.046 & 0.039 \\
 1.000 ($\sigma_e$) & 1.001 & 0.016 & 0.013 & 1.002 & 0.018 & 0.013 & 1.048 & 0.028 & 0.049 \\
  1.000 ($\sigma_1$) & 1.017 & 0.104 & 0.078 & 1.033 & 0.116 & 0.082 & 0.779 & 0.093 & 0.221 \\
  0.250 ($\sigma_2$) & 0.246 & 0.025 & 0.030 & 0.250 & 0.038 & 0.030 & 0.184 & 0.019 & 0.067 \\
\hline
& \multicolumn{9}{c}{$n_{ind} = 200, n_{rep} = 20$} \\
\hline
-2.000 ($\beta_0$)  & -1.989 & 0.042 & 0.069 & -2.002 & 0.074 & 0.057  & -1.937 & 0.100 & 0.112 \\
 1.500 ($\beta_1$)  & 1.500 & 0.034 & 0.033 & 1.500 & 0.040 & 0.033    & 1.497 & 0.041 & 0.034 \\
1.000 ($\sigma_e$) & 1.003 & 0.011 & 0.012 & 1.001 & 0.012 & 0.009 & 1.052 & 0.028 & 0.052 \\
1.000 ($\sigma_1$) & 1.017 & 0.104 & 0.078 & 1.033 & 0.116 & 0.082 & 0.741 & 0.113 & 0.259 \\
0.250 ($\sigma_2$) & 0.254 & 0.026 & 0.023 & 0.257 & 0.032 & 0.023 & 0.191 & 0.016 & 0.059 \\
\hline
\end{tabular}
\label{resultados_rho0_sim3}
\end{table}

\begin{table}
\centering
    \caption{True values of the parameters, average posterior mean (Mean) and average standard deviation (SD), and MAE for all parameters of the model (\ref{model1_simu}) in different scenarios with $\rho = 0.7$.}
\begin{tabular}{r|rrr|rrr|rrr}
\hline
&  \multicolumn{9}{c}{$n_{ind} = 100, n_{rep} = 10$} \\
\hline
& \multicolumn{3}{c|}{VB + Wavelets} & \multicolumn{3}{c|}{NUTS} & \multicolumn{3}{c}{ADVI} \\ 
 True values & Mean & SD & MAE & Mean & SD & MAE & Mean & SD & MAE \\
\hline
 -2.000 ($\beta_0$) & -2.003 & 0.073 & 0.085 & -2.003 & 0.106 & 0.085 & -1.970 & 0.117 & 0.112   \\
  1.500 ($\beta_1$)  & 1.495 & 0.051 & 0.050 & 1.495 & 0.062 & 0.049  & 1.494 & 0.066 & 0.050  \\
  1.000 ($\sigma_e$) & 0.999 & 0.023 & 0.023 & 0.998 & 0.025 & 0.023  & 1.033 & 0.031 & 0.042 \\
  1.000 ($\sigma_1$)& 1.018 & 0.124 & 0.118 & 1.039 & 0.165 & 0.124 & 0.830 & 0.115 & 0.176 \\
  0.250 ($\sigma_2$) & 0.261 & 0.032 & 0.044 & 0.265 & 0.056 & 0.046 & 0.201 & 0.037 & 0.054 \\
  0.350 ($\sigma_{12}$) & 0.341 & 0.038 & 0.063 & 0.345 & 0.075 & 0.062 & 0.269 & 0.047 & 0.089 \\
\hline
&  \multicolumn{9}{c}{$n_{ind} = 100, n_{rep} = 20$} \\
\hline
-2.000 ($\beta_0$)   & -2.001 & 0.061 & 0.079 & -2.001 & 0.104 & 0.082 & -1.934 & 0.127 & 0.136  \\
1.500  ($\beta_1$)  & 1.498 & 0.042 & 0.048 & 1.497 & 0.056 & 0.049    & 1.499 & 0.059 & 0.047  \\
1.000  ($\sigma_e$) & 1.002 & 0.016 & 0.017 & 1.001 & 0.017 & 0.015 & 1.041 & 0.027 & 0.047 \\
1.000  ($\sigma_1$) & 1.024 & 0.121 & 0.104 & 1.043 & 0.158 & 0.105 & 0.812 & 0.125 & 0.192  \\
0.250  ($\sigma_2$) & 0.261 & 0.031 & 0.036 & 0.266 & 0.046 & 0.037 & 0.201 & 0.029 & 0.053 \\
0.350  ($\sigma_{12}$) & 0.352 & 0.037 & 0.049 & 0.357 & 0.070 & 0.050 & 0.261 & 0.041 & 0.093 \\
\hline
& \multicolumn{9}{c}{$n_{ind} = 200, n_{rep} = 10$} \\
\hline
-2.000 ($\beta_0$)   & -2.008 & 0.044 & 0.061 & -2.006 & 0.075 & 0.061 & -2.006 & 0.072 & 0.061 \\  
1.500  ($\beta_1$)  & 1.495 & 0.035 & 0.042 & 1.496 & 0.043 & 0.041    & 1.495 & 0.043 & 0.041 \\ 
1.000  ($\sigma_e$) & 0.999 & 0.016 & 0.016 & 0.998 & 0.018 & 0.014  & 1.026 & 0.022 & 0.027\\
1.000  ($\sigma_1$) & 1.027 & 0.088 & 0.089 & 1.035 & 0.115 & 0.091 & 0.776 & 0.066 & 0.224 \\
0.250  ($\sigma_2$) & 0.260 & 0.022 & 0.031 & 0.259 & 0.038 & 0.030 & 0.188 & 0.021 & 0.062 \\
0.350  ($\sigma_{12}$) & 0.348 & 0.027 & 0.045 & 0.353 & 0.053 & 0.045 & 0.266 & 0.031 & 0.084  \\
\hline
&  \multicolumn{9}{c}{$n_{ind} = 200, n_{rep} = 20$} \\
\hline
($\beta_0$)  -2.000 & -2.005 & 0.035 & 0.063 & -2.004 & 0.073 & 0.062& -2.000 & 0.067 & 0.062 \\ 
($\beta_1$)  1.500 & 1.497 & 0.028 & 0.039 & 1.496 & 0.039 & 0.038  & 1.495 & 0.040 & 0.038 \\ 
($\sigma_e$) 1.000 & 0.999 & 0.011 & 0.011 & 1.000 & 0.012 & 0.010 & 1.015 & 0.014 & 0.016 \\
($\sigma_1$) 1.000 & 1.019 & 0.086 & 0.081 & 1.026 & 0.109 & 0.079 & 0.723 & 0.053 & 0.277  \\
($\sigma_2$) 0.250 & 0.254 & 0.022 & 0.023 & 0.256 & 0.031 & 0.023 & 0.186 & 0.016 & 0.064 \\
($\sigma_{12}$) 0.350 & 0.350 & 0.026 & 0.039 & 0.353 & 0.048 & 0.038 & 0.246 & 0.025 & 0.104 \\
\hline
\end{tabular}
\label{resultados_rho07_sim3}
\end{table}

\begin{figure}
    \centering
    \includegraphics[width=0.75\linewidth]{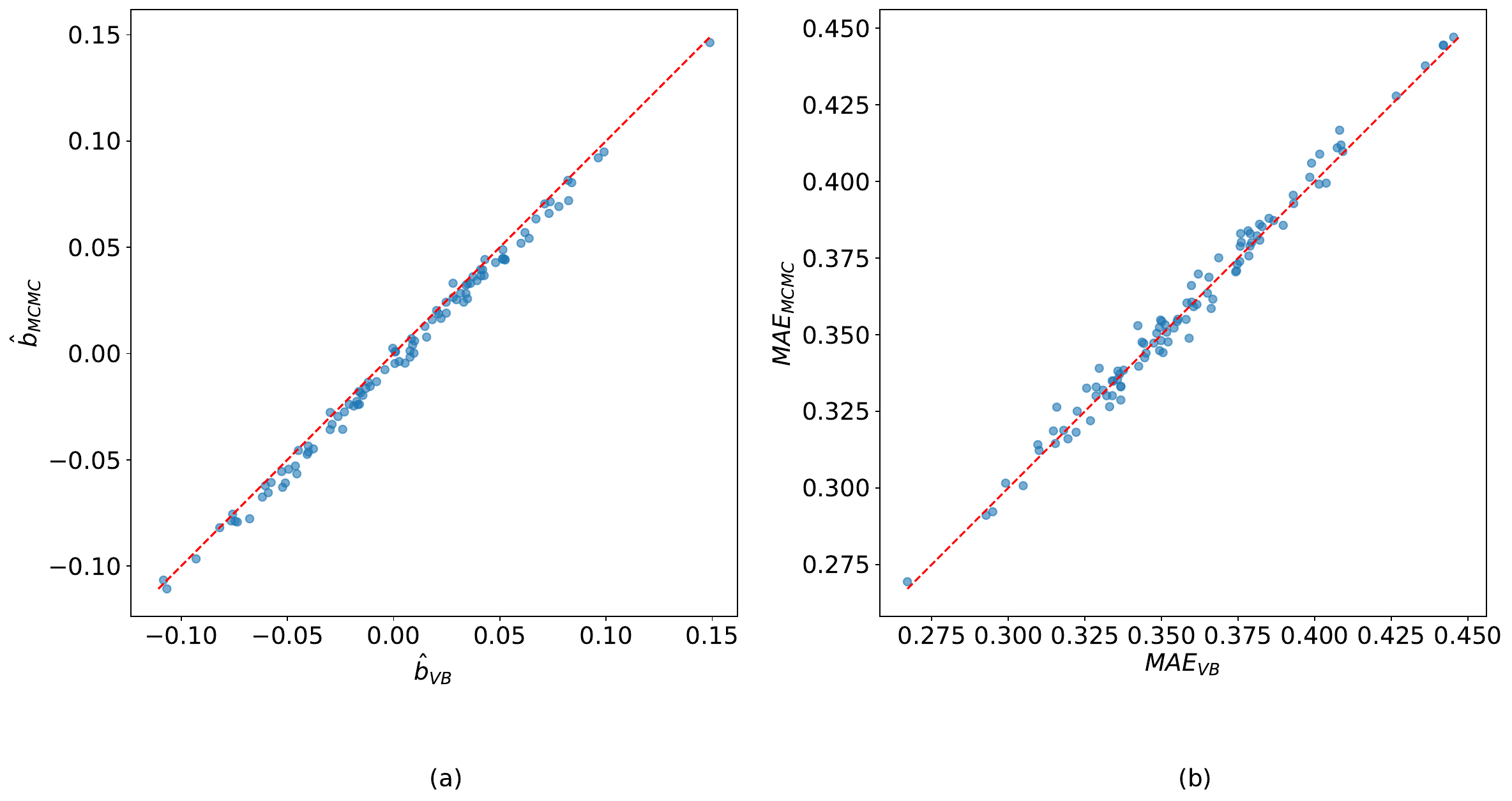}
        \caption{Scatter plot in (a) shows the comparison between MCMC estimates ($\hat{b}_{MCMC}$) and VB estimates $\hat{b}_{VB}$. Scatter plot in (b) shows the comparison between MAE of MCMC estimates ($\hat{b}_{MCMC}$) and MAE of VB estimates $\hat{b}_{VB}$. Both for scenario $\rho = 0$, $n_{ind} = 100, n_{rep} = 10$. }
    \label{scatter1_simu3}
\end{figure}

\begin{figure}
    \centering
    \includegraphics[width=0.75\linewidth]{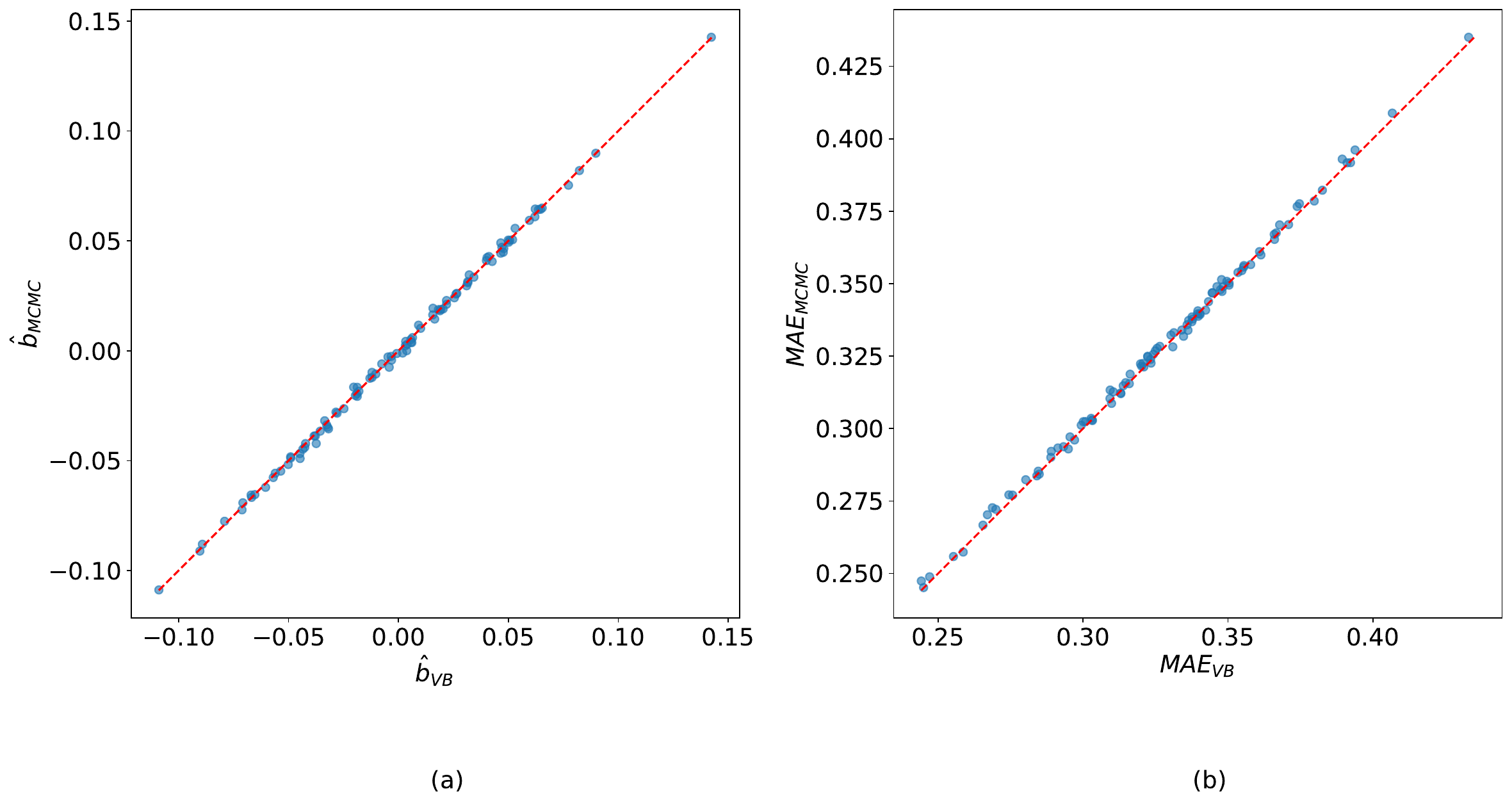}
    \caption{Scatter plot in (a) shows the comparison between MCMC estimates ($\hat{b}_{MCMC}$) and VB estimates $\hat{b}_{VB}$. Scatter plot in (b) shows the comparison between MAE of MCMC estimates ($\hat{b}_{MCMC}$) and MAE of VB estimates $\hat{b}_{VB}$. Both for scenario $\rho = 0$, $n_{ind} = 100, n_{rep} = 20$. }
    \label{scatter2_simu3}
\end{figure}

In Figures \ref{scatter1_simu3} and \ref{scatter2_simu3}, we illustrate the similarity between the posterior mean estimates of $\boldsymbol{b}$ obtained from MCMC and VB, as well as the MAE comparison between these two approaches.

\section{Applications}

\subsection{Hierarchical logistic regression}

In this section we explore a hierarchical model first presented by \cite{gelman2007data} who proposed to estimate state-level opinions from national polls, while simultaneously correcting for non-response, for any survey response of interest. The model was chosen for this application because involves many random effects and hyperparameters, and it was also used as application in \cite{kucukelbir2017automatic}.

The model is a regression for the individual response $y$ given demographics and state. In the dataset we have sex (male or female), ethnicity (African American or other), age (4 categories), education (4 categories), 51 states, 5 regions of the country (Northeast, Midwest, South, West, and D.C.), and a measure of previous Republican vote in the state. 

The outcome for the $i-$th individual on the survey, $y_i$ is labeled as 1 if supports the Republican candidate and 0 if supports Democrat and model them as independent given the demographics characteristics and state. So, the outcome of the model is the probability that a respondent prefers the Republican candidate for president estimated by a logistic regression model from a set of seven CBS News polls conducted during the week before the 1988 presidential election. We can write the model in the hierarchical form using the indexes $j,k,l, m$ for state, age category, education category, and region:

\begin{align}\label{model_ap1}
    P(y_i =1) & = \text{logit}^{-1}( \beta_0 + \beta_1 \cdot \text{female}_i + \beta_2 \cdot \text{black}_i + \\ & \beta_3 \cdot \text{female}_i \cdot \text{black}_i + \alpha_{k[i]}^{\text{age}} + \alpha_{l[i]}^{\text{edu}} + \alpha_{k[i],l[i]}^{\text{age.edu}} + \alpha_{j[i]}^{\text{state}}) \\ \nonumber
    \alpha_j^{\text{state}} & \sim N(\alpha_{m[j]}^{\text{region}} + \beta_4 \cdot \text{v.prev}_j, \sigma^2_{\text{state}}) \\ \nonumber
    \alpha_k^{\text{age}} & \sim N(0, \sigma^2_{\text{age}})\\\nonumber
    \alpha_k^{\text{edu}} & \sim N(0, \sigma^2_{\text{edu}})\\\nonumber
    \alpha_k^{\text{age.edu}} & \sim N(0, \sigma^2_{\text{age.edu}})\\\nonumber
    \alpha_k^{\text{region}} & \sim N(0, \sigma^2_{\text{region}})
\end{align}

\noindent for $k =1, \hdots, 4$, $l =1, \hdots, 4$, $m =1, \hdots, 5$, $j = 1, \hdots, 51$, and $i = 1, \hdots, 2015.$ The authors in \cite{gelman2007data} defined this model interested in the estimate proportion of Republican votes in many different categories. As consequence, we have few or no data in many of them, however, this is not a problem for this model. This model is challenging for bayesian fitting procedure, besides the large number of categories, there is also a moderate number of hyperparameters. 

We considered the Wavelets-copula variational family defined in Section \ref{section_copula_family} considering a Gaussian Copula. The optimization of the variational objective defined in (\ref{objective}) was performed using 2000 iterations of the RMSprop algorithm, as implemented in PyTorch, with learning rate set to 0.01, smoothing constant $\alpha = 0.5$, and momentum parameter equal to 0.9. The number of Monte Carlo samples used was 50. We benchmarked our proposal with the NUTS algorithm, and ADVI both implemented in Stan. For the NUTS we run on chain of size 4000, and discarded 1000 as burn-in, this chain took 231.0 seconds to simulate. For the ADVI, we considered the full rank approach with 300000 iteration which took 82.5 seconds. Our approach took 7.6 seconds

Table \ref{fixed_effects_m1} shows some posterior quantities, mean, standard deviation and quantiles, for the fixed effects and hyperparameters of the model (\ref{model_ap1}).

\begin{table}[htpb]
\centering
\scriptsize
    \caption{Posterior mean (Mean), standard deviation, 0.025 quantile, and 0.975 quantile estimates for the Variational Bayesis with Wavelet Copula family, NUTS, and ADVI for all parameters of the model (\ref{model_ap1}).}
\begin{tabular}{l|rrrr|rrrr|rrrr}
\hline
& \multicolumn{4}{c|}{VB + Wavelets} & \multicolumn{4}{c|}{NUTS} & \multicolumn{4}{c}{ADVI}\\  
 & Mean & SD & $q_{0.025}$ & $q_{0.975}$ & Mean & SD & $q_{0.025}$ & $q_{0.975}$ & Mean & SD & $q_{0.025}$ & $q_{0.975}$ \\
\hline
$\beta_0$ &-1.913 & 0.780 & -3.439 & -0.381 & -1.891 & 0.873 & -3.657 & -0.198 & 0.601 & 0.699 & -0.801 & 1.897 \\
$\beta_1$ & -1.835 & 0.336 & -2.518 & -1.195 & -1.823 & 0.349 & -2.524 & -1.169 & -1.906 & 0.382 & -2.631 & -1.166 \\
$\beta_2$ & -0.101 & 0.102 & -0.294 & 0.095 & -0.112 & 0.103 & -0.312 & 0.086 & -0.126 & 0.115 & -0.347 & 0.108 \\
$\beta_3$ & 0.094 & 0.444 & -0.787 & 0.957 & 0.087 & 0.437 & -0.752 & 0.964 & 0.109 & 0.485 & -0.836 & 1.099 \\
$\beta_4$ & 4.194 & 1.387 & 1.521 & 6.877 & 4.123 & 1.490 & 1.242 & 7.086 & 1.157 & 1.135 & -0.971 & 3.288\\
$\sigma$ & 0.185 & 0.110 & 0.075 & 0.491 & 0.171 & 0.171 & 0.010 & 0.630 & 0.327 & 0.157 & 0.121 & 0.658 \\
$\sigma_{age}$ & 0.239 & 0.157 & 0.090 & 0.644 & 0.219 & 0.203 & 0.015 & 0.766 & 0.907 & 0.588 & 0.227 & 2.358 \\
$\sigma_{edu}$ & 0.249 & 0.029 & 0.200 & 0.312 & 0.192 & 0.088 & 0.039 & 0.372 & 0.373 & 0.047 & 0.287 & 0.473 \\
$\sigma_{state}$ & 0.494 & 0.218 & 0.240 & 1.089 & 0.521 & 0.296 & 0.184 & 1.304 & 0.579 & 0.242 & 0.233 & 1.156 \\
$\sigma_{region}$ & 0.189 & 0.046 & 0.122 & 0.301 & 0.159 & 0.101 & 0.020 & 0.387 & 0.336 & 0.077 & 0.211 & 0.497 \\
\hline
\end{tabular}
    \label{fixed_effects_m1}
\end{table}

There is strong agreement between VI with Wavelets-copula family and NUTS across most parameters presented in Table \ref{fixed_effects_m1}, both in terms of posterior means and uncertainty quantification, indicating that the proposed variational approach closely approximates the MCMC-based inference. In contrast, ADVI exhibits noticeable discrepancies for several parameters, suggesting a less accurate approximation in this setting.

The coefficient $\beta_4$ shows relatively large posterior uncertainty under both VI with Wavelets and NUTS. However, ADVI substantially underestimates this parameter and yields a posterior distribution that is clearly shifted relative to the other two methods. The strongest discrepancy presented by the ADVI was the results for the hyper parameters, posterior mean for all these distributions disagree with those presented by VI with Wavelets and NUTS.

\begin{figure}[htpb]
    \centering
    \includegraphics[width=0.5\linewidth]{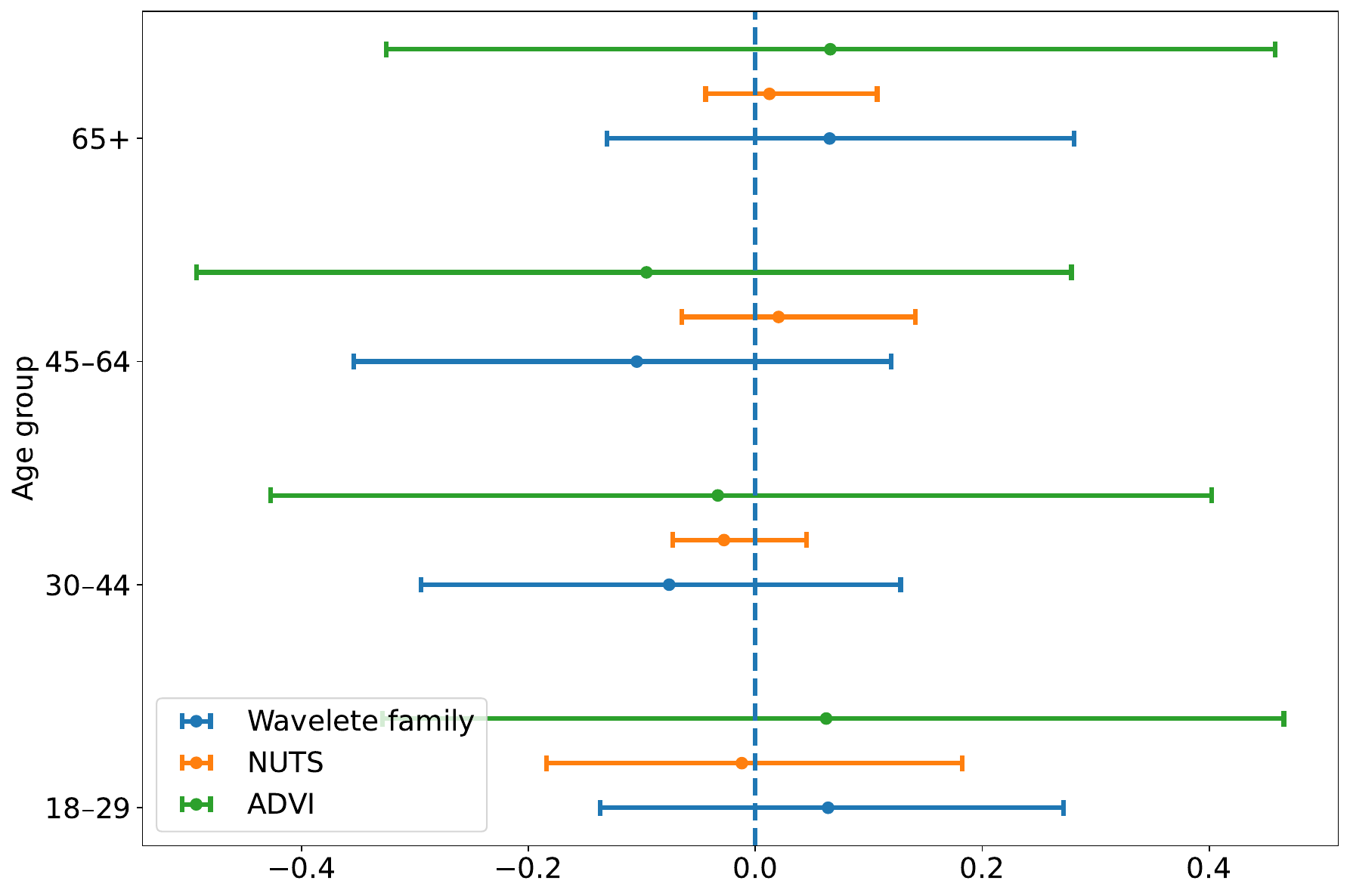}
    \caption{Posterior mean estimates (dots) with HPD intervals for all age groups separeted by methodology.}
    \label{age_groups}
\end{figure}

The Figure \ref{age_groups} illustrates the posterior distribution of the age group effects (18–29, 30–44, 45–64, and 65+), $\alpha_k$ posterior distributions for $k = 1, \hdots, 4,$ obtained using the Wavelet Copula variational family, NUTS, and ADVI. The lines represent the High Density Posterior Intervals (HPD), and the dots the posterior means. Overall, all Variational methods (Wavelets family and ADVI) overestimates the posterior variance compared to the MCMC method (NUTS), but presented posterior mean estimates close to the NUTS approach. 

\subsection{Bayesian Conditional Transformation Model}

In this section, we employ the proposed Variational family in a distinct model from the first application. We considered the Bayesian Conditional Transformation Model (BCTM) \cite{carlan2023bayesian}. Precisely, we reproduce an application using this model presented in \cite{Giovanni25}. The model considered has many non-linear terms and hyperparameters, and it is a challenge for the proposed methodology. \cite{carlan2023bayesian} proposed a NUTS sampler for the BCTM, despite the precision of the algorithm, it takes much time to converge. \cite{Giovanni25} proposed a new INLA based algorithm for the model. In fact, the proposed in much faster the NUTS approach, but can suffer from a moderate/high number of hyperprameters.

The BCTM aims to estimate the conditional distribution function of a response variable \( Y \), given an observed explanatory vector \( \boldsymbol{X} = \boldsymbol{x} \), denoted as \( F_{Y|\boldsymbol{X} = \boldsymbol{x}} \). The model write this response variable as a transformation such that the transformed variable follows a known probability distribution called baseline distribution. For simplicity, the baseline is often assumed to be a normal distribution. This is defined as follows:

\begin{equation}\label{conditional_distribution}
     F_{Y|\boldsymbol{X} = \boldsymbol{x}}(y) = P(Y \leq y|\boldsymbol{X} = \boldsymbol{x}) = P(h(Y|\boldsymbol{x}) \leq h(y|\boldsymbol{x})|\boldsymbol{X} = \boldsymbol{x}) = F_Z(h(y|\boldsymbol{x})),
\end{equation}

\noindent where \( F_Z \) is the cdf of standard normal distribution. The crucial aspect of the CTM class is the transformation function, and the construction of \( h(y|\boldsymbol{x}) \) must account for all the characteristics and objectives of the proposed model. It is convenient to decompose the transformation function \( h \) additively into \( J \in \mathbb{N}\) partial transformation functions $h_j$ such that

\begin{equation}\label{decomp_h}
    h(y|\boldsymbol{x}) = \sum_{j = 1}^J h_j(y|\boldsymbol{x}).
\end{equation}

The transformation functions \( h_j(y|\boldsymbol{x}) \) in (\ref{decomp_h}) can range from low-parameterized forms to complex structures with many parameters. Besides that, it is clear that the transformation function $h(y|\boldsymbol{x})$ should be monotone in $y$ because of the equation in (\ref{conditional_distribution}). 

The context of this application comes from examining a time series of reported thefts. This study aims to describe temporal patterns in vehicle theft incidents in São Paulo. All vehicle theft incidents (VTIs), involving only robbery, recorded by the police in the city between January 1, 2017, and December 31, 2024, were obtained from the website of the São Paulo State Secretariat of Public Security (\url{https://www.ssp.sp.gov.br/estatistica/consultas}). The raw data consists of detailed records of each reported VTI in the city of São Paulo, including the specific date (year, month and day) and the hour of occurrence. Under-reporting is not a concern, as a police report is mandatory for any car insurance claim in Brazil. Moreover, the possibility of the vehicle being recovered by the owner is more likely with an official police record.

For the purposes of this analysis, we aggregate the number of VTIs into pre-defined time intervals. This aggregation make it easier to identify the broader temporal patterns and trends that might not be apparent when analyzing each hour and day of a time series. We define the count response variable $Y$ as the number of VTIs within each specific time interval. The pre-defined time intervals accomplished, Year, Month, Day of the week, and interval of the day. The specific intervals of the day are labeled as: ``night am" [00:00 - 04:00], ``morning" [04:00 - 08:00], ``day am" [08:00 - 12:00], ``day pm" [12:00 - 16:00], ``pre night" [16:00 - 20:00], and ``night" [20:00 - 24:00]. The annual effect, calculated as the difference between each year and the base year (2017) plus one; the weekly effect, measured as the difference between each day of the week and Sunday plus one; and the monthly effect, defined as the difference between each month and January plus one. The vector of covariates $\boldsymbol{X}$ encapsulates all these covariates. The data was split into training and testing sets, with 90\% of the data used for training (a total of 4,232 observations) and the remaining 10\% reserved for testing (a total of 472 observations).


We considered the following BCTM proposed by \cite{Giovanni25}:

\begin{align}\label{bctm}
   P(Y_i \leq y_i| \boldsymbol{X}_i = \boldsymbol{x}_i) = & \Phi(\boldsymbol{a}_{20}(\lfloor y_i \rfloor)^T \boldsymbol{\gamma}_1 + \beta_0 + \boldsymbol{a}_5(Year_i)^T\boldsymbol{\gamma}_2 + \boldsymbol{a}_5(Month_i)^T\boldsymbol{\gamma}_3 \\
    & +  (\boldsymbol{a}_5(Day_i)^T \otimes (\textrm{Period}_i)^T) \boldsymbol{\gamma}_4 + \textrm{Period}_i \times \boldsymbol{\gamma}_5). \nonumber
\end{align}

\noindent where $\boldsymbol{a}_{20}(y)$ is a monotonic Bernstein polynomial of degree 20, $\lfloor y \rfloor$ is the floor dunction because the response variable is the count type, and $\boldsymbol{a}_5(Year)$, and $\boldsymbol{a}_5(Month)$ represent Bernstein polynomials of degree 5 for year, and month effects. The term $(\boldsymbol{a}_5(Day)^T \otimes (\textrm{Period})^T)$ represents a interaction between day effects (in terms of a Bernstein polynomial) and period of the day (a categorical variable).

In (\ref{bctm}) we have that the vectors $\boldsymbol{\gamma}_2, \boldsymbol{\gamma}_3, \boldsymbol{\gamma}_4,$ and $\boldsymbol{\gamma}_5$ are real valued. However, the $\boldsymbol{\gamma}_1 = \boldsymbol{\Sigma} \boldsymbol{\Tilde{\beta}}_1$ with $\boldsymbol{\tilde{\beta}}_1 = (\beta_{1,1}, \exp(\beta_{1,2}), \hdots, \exp(\beta_{1,20}))^{\top}$, and $\boldsymbol{\Sigma}$ is a lower block triangular matrix of dimension $20$, meaning that the elements above the diagonal are zero and the remaining elements are 1. We set the priors $\boldsymbol{\beta}_1 \sim N(\boldsymbol{0}, \frac{1}{\tau_1}\boldsymbol{K}_1)$, and $\boldsymbol{\gamma}_j \sim N(\boldsymbol{0}, \frac{1}{\tau_j}\boldsymbol{K}_1)$, for $j = 2, 3$, and $\boldsymbol{\gamma}_4 \sim N(0, \frac{1}{\tau_4}(\boldsymbol{K}_1 \otimes I_6))$. For $\boldsymbol{\gamma}_5$ we set independent normal priors with mean zero and variance 100.

This model has a total of 36 parameters and 4 hyperparameters. 
We considered the Wavelet Gaussian copula family with Adam optimizer. The algorithm \ref{alg1} took 9.74 seconds with 2000 epochs and 50 Monte Carlo samples size.

\begin{figure}[h]
    \centering
        \centering
        \includegraphics[scale = 0.45]{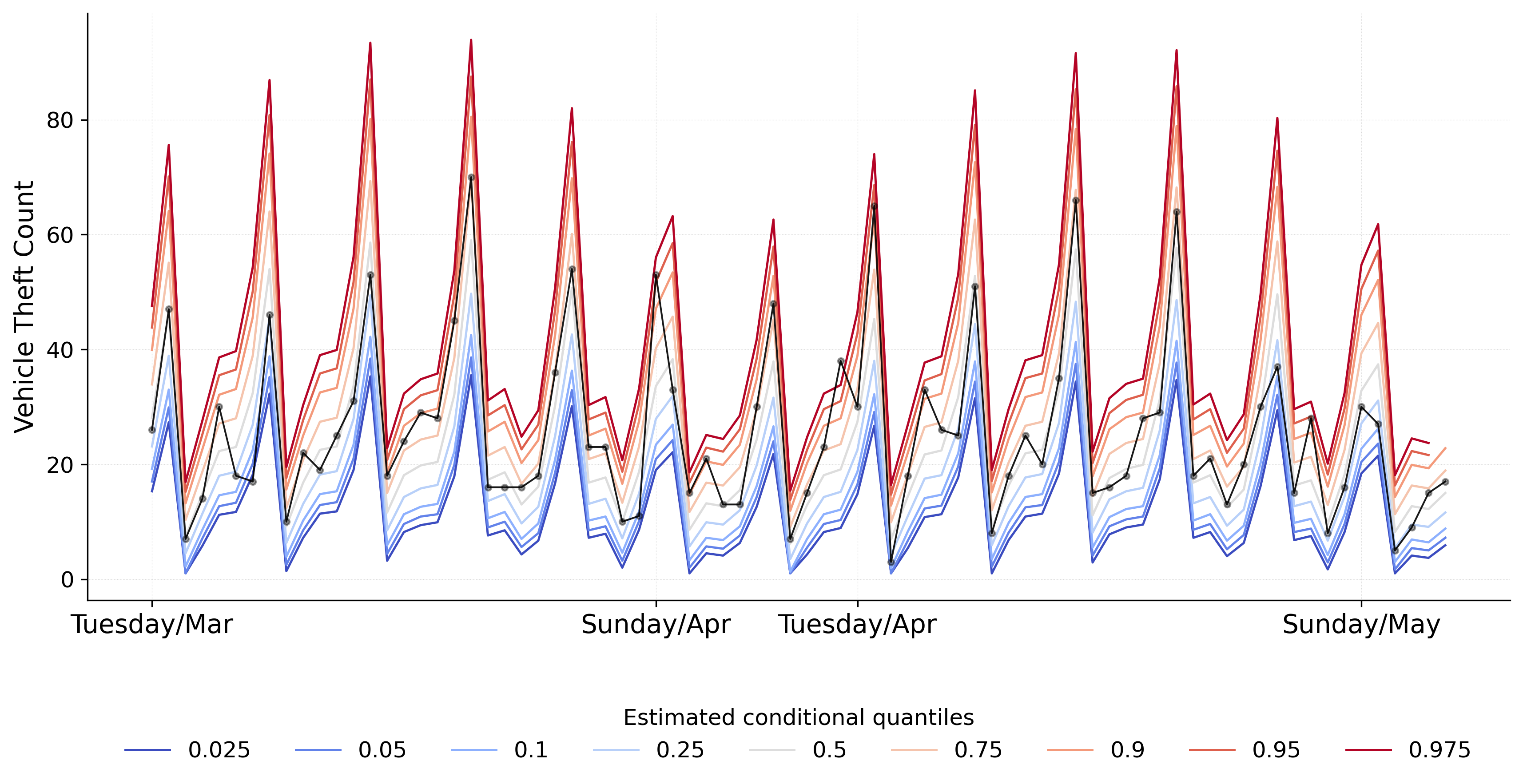}
        \caption{Estimated quantiles of the conditional distribution from BCTM Model of the first 700 observations in the test set given the explanatory variables. The black line is the observed count on the test set.}
        \label{predictions_m3}
\end{figure}

We obtained the posterior distributions of the $\boldsymbol{\gamma}$'s parameters and the hyperparameters. From that, we estimated the conditional distributions, $F_{Y|\boldsymbol{X_i}}$ from the observations in the dataset. Figure \ref{predictions_m3} displays the estimated conditional quantiles for the first 90 observations of the test dataset. The conditional quantile predictions suggest an interval estimation for different probabilities. For instance, the 0.5 quantile (median) indicates the central tendency of the predictions, while higher and lower quantiles provide bounds for uncertainty. We evaluated the MAE (Mean Absolute Error) on the test set. The MAE is calculated in the test data as $\sum_{i = 1}^n |y_i - E(Y_i| \boldsymbol{X}_i = \boldsymbol{x}_i)|/n$, where $E(Y_i| \boldsymbol{X}_i = \boldsymbol{x}_i)$ represents the expected value of the i-th conditional random variable $Y_i$ given the explanatory variables $\boldsymbol{x}_i$. We obtained 5.462, very similar from that obtained in \cite{Giovanni25} (5.460)

\begin{figure}[h]
     \centering
         \includegraphics[scale = 0.25]{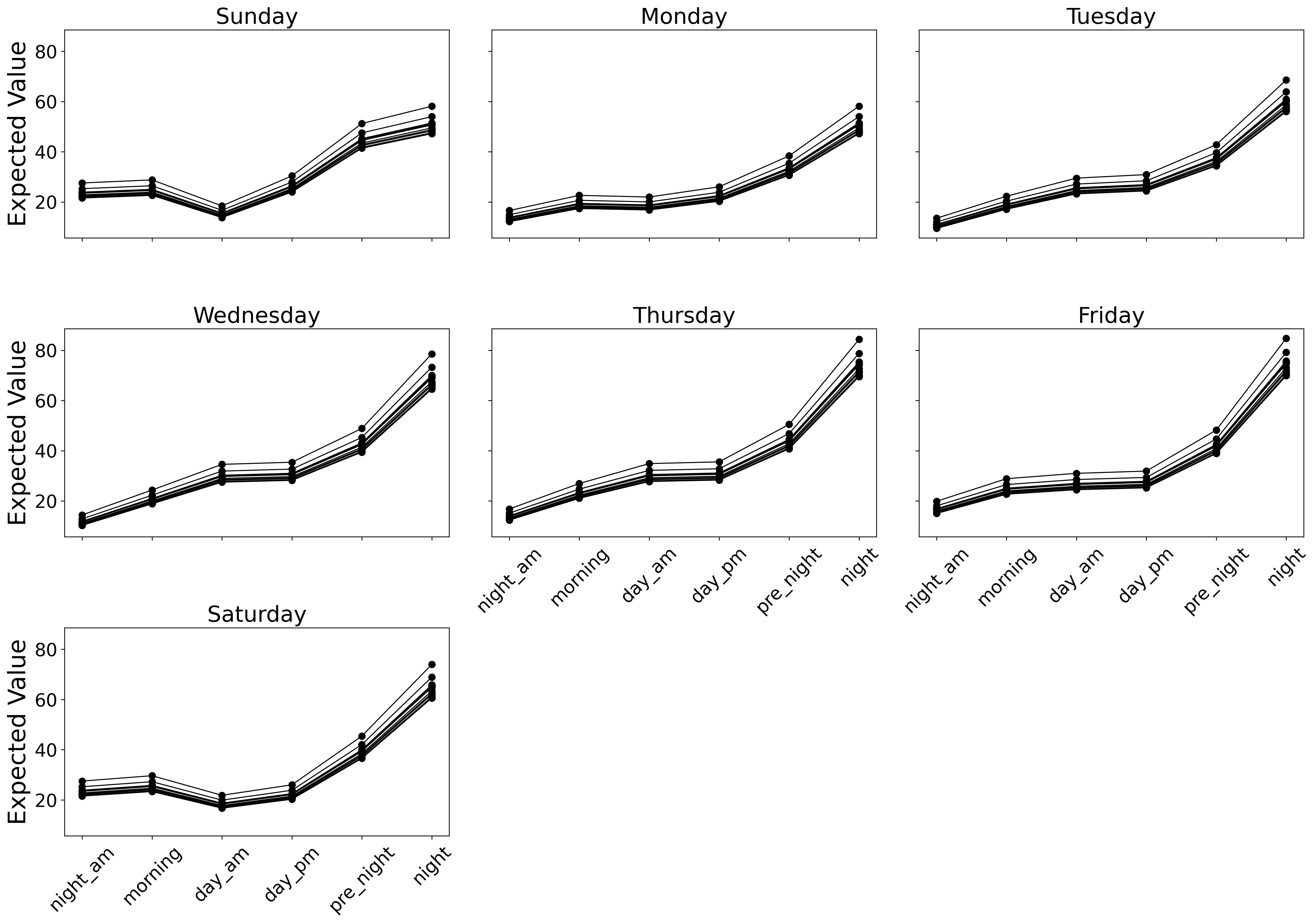}
         \caption{Expected value of $Y|\boldsymbol{x}$ given $year = 2019$. Each line represent a different month.}
         \label{expected_2019}
\end{figure}

Figure \ref{expected_2019} displays the calculated conditional expected value for the year 2019. The x-axis represents the period of the day and y-axis represents the expected value. These measures are evaluated fixing the period of the Day (x-axis), the day of the week (top of the graph), the year as 2019, and each line represents a different month (12 different lines). The Figure \ref{expected_2019} illustrates how the conditional expected value of vehicle theft varies throughout the day. The highest expected number of vehicle thefts occurs at night (20:00 to 24:00) for every day of the week. Additionally, the expected number of vehicle thefts is lower on weekends compared to the rest of the week, particularly in the afternoon.

\clearpage

\end{document}